\documentclass[twoside,twocolumn,9pt]{article}
\usepackage{extsizes}
\usepackage[super,sort&compress,comma]{natbib} 
\usepackage[version=3]{mhchem}
\usepackage[left=1.5cm, right=1.5cm, top=1.785cm, bottom=2.0cm]{geometry}
\usepackage{balance}
\usepackage{mathptmx}
\usepackage{sectsty}
\usepackage{graphicx} 
\usepackage{lastpage}
\usepackage[format=plain,justification=justified,singlelinecheck=false,font={stretch=1.125,small,sf},labelfont=bf,labelsep=space]{caption}
\usepackage{float}
\usepackage{fancyhdr}
\usepackage{fnpos}
\usepackage[english]{babel}
\addto{\captionsenglish}{%
  
}
\usepackage{array}
\usepackage{droidsans}
\usepackage{charter}
\usepackage[T1]{fontenc}
\usepackage[usenames,dvipsnames]{xcolor}
\usepackage{setspace}
\usepackage[compact]{titlesec}
\usepackage{hyperref}

\usepackage{epstopdf}

\definecolor{cream}{RGB}{222,217,201}

\begin{document}

\pagestyle{fancy}
\thispagestyle{plain}
\fancypagestyle{plain}{
\renewcommand{\headrulewidth}{0pt}
}

\makeFNbottom
\makeatletter
\renewcommand\LARGE{\@setfontsize\LARGE{15pt}{17}}
\renewcommand\Large{\@setfontsize\Large{12pt}{14}}
\renewcommand\large{\@setfontsize\large{10pt}{12}}
\renewcommand\footnotesize{\@setfontsize\footnotesize{7pt}{10}}
\makeatother

\renewcommand{\thefootnote}{\fnsymbol{footnote}}
\renewcommand\footnoterule{\vspace*{1pt}%
\color{cream}\hrule width 3.5in height 0.4pt \color{black}\vspace*{5pt}} 
\setcounter{secnumdepth}{5}

\makeatletter 
\renewcommand\@biblabel[1]{#1}            
\renewcommand\@makefntext[1]%
{\noindent\makebox[0pt][r]{\@thefnmark\,}#1}
\makeatother 
\renewcommand{\figurename}{\small{Fig.}~}
\sectionfont{\sffamily\Large}
\subsectionfont{\normalsize}
\subsubsectionfont{\bf}
\setstretch{1.125} 
\setlength{\skip\footins}{0.8cm}
\setlength{\footnotesep}{0.25cm}
\setlength{\jot}{10pt}
\titlespacing*{\section}{0pt}{4pt}{4pt}
\titlespacing*{\subsection}{0pt}{15pt}{1pt}

\fancyfoot{}
\fancyfoot[LO,RE]{\vspace{-7.1pt}\includegraphics[height=9pt]{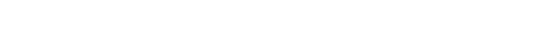}}
\fancyfoot[CO]{\vspace{-7.1pt}\hspace{13.2cm}\includegraphics{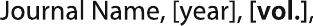}}
\fancyfoot[CE]{\vspace{-7.2pt}\hspace{-14.2cm}\includegraphics{head_foot/RF}}
\fancyfoot[RO]{\footnotesize{\sffamily{1--\pageref{LastPage} ~\textbar  \hspace{2pt}\thepage}}}
\fancyfoot[LE]{\footnotesize{\sffamily{\thepage~\textbar\hspace{3.45cm} 1--\pageref{LastPage}}}}
\fancyhead{}
\renewcommand{\headrulewidth}{0pt} 
\renewcommand{\footrulewidth}{0pt}
\setlength{\arrayrulewidth}{1pt}
\setlength{\columnsep}{6.5mm}
\setlength\bibsep{1pt}

\makeatletter 
\newlength{\figrulesep} 
\setlength{\figrulesep}{0.5\textfloatsep} 

\newcommand{\topfigrule}{\vspace*{-1pt}%
\noindent{\color{cream}\rule[-\figrulesep]{\columnwidth}{1.5pt}} }

\newcommand{\botfigrule}{\vspace*{-2pt}%
\noindent{\color{cream}\rule[\figrulesep]{\columnwidth}{1.5pt}} }

\newcommand{\dblfigrule}{\vspace*{-1pt}%
\noindent{\color{cream}\rule[-\figrulesep]{\textwidth}{1.5pt}} }

\makeatother

\twocolumn[
  \begin{@twocolumnfalse}
{\includegraphics[height=30pt]{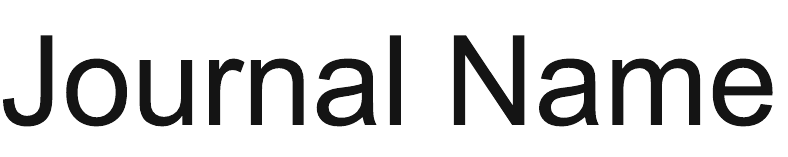}\hfill\raisebox{0pt}[0pt][0pt]{\includegraphics[height=55pt]{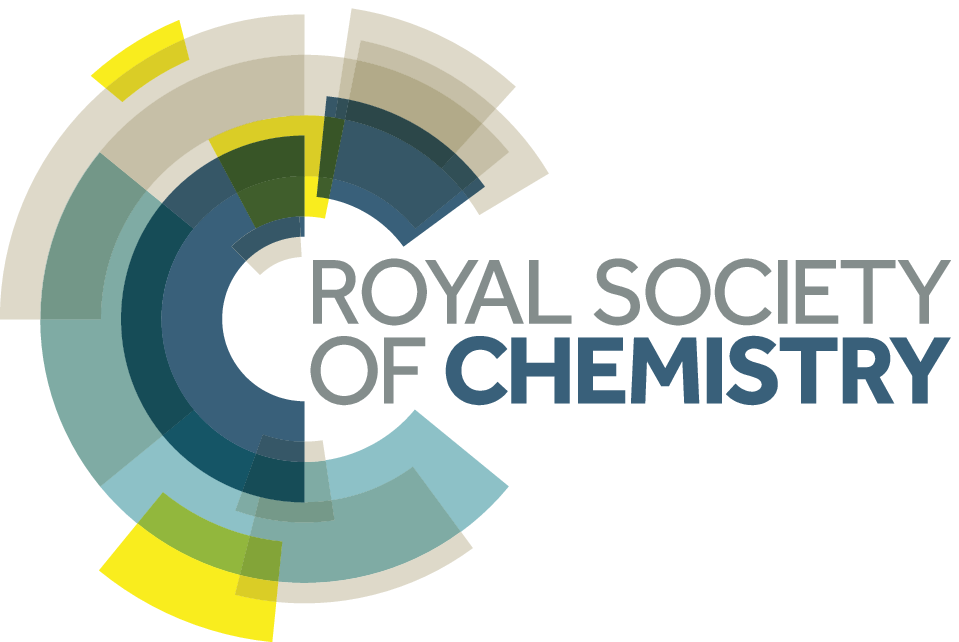}}\\[1ex]
\includegraphics[width=18.5cm]{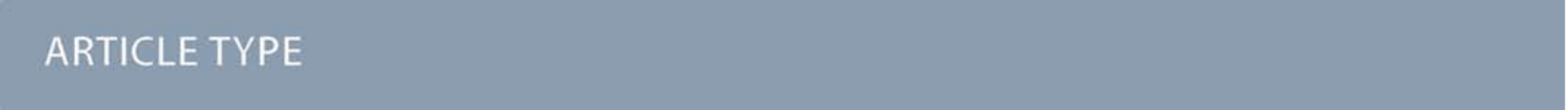}}\par
\vspace{1em}
\sffamily
\begin{tabular}{m{4.5cm} p{13.5cm} }

\includegraphics{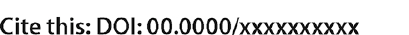} & \noindent\LARGE{\textbf{A quantitative model for a nanoscale switch accurately predicts thermal actuation behavior
$^\dag$}} \\
\vspace{0.3cm} & \vspace{0.3cm} \\

 & \noindent\large{Kyle Crocker,\textit{$^{a}$} Joshua Johnson,\textit{$^{bc}$} Wolfgang Pfeifer\textit{$^{ad}$}, Carlos Castro\textit{$^{d}$}, and Ralf Bundschuh$^{\ast}$\textit{$^{abefg}$}} \\

\includegraphics{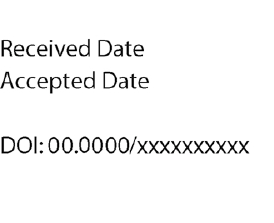} & \noindent\normalsize{Manipulation of temperature can be used to actuate DNA origami nano-hinges containing gold nanoparticles. We develop a physical model of this system that uses partition function analysis of the interaction between the nano-hinge and nanoparticle to predict the probability that the nano-hinge is open at a given temperature. The model agrees well with experimental data and predicts experimental conditions that allow the actuation temperature of the nano-hinge to be tuned over a range of temperatures from $30$${}^{\circ}\mathrm{C}$ to $45$${}^{\circ}\mathrm{C}$. Additionally, the model reveals surprising physical constraints on the system. This combination of physical insight and predictive potential is likely to inform future designs that integrate nanoparticles into dynamic DNA origami structures. Furthermore, our modeling approach could be expanded to consider the incorporation, stability, and actuation of other types of functional elements or actuation mechanisms integrated into nucleic acid devices.}

\end{tabular}

 \end{@twocolumnfalse} \vspace{0.6cm}

  ]

\renewcommand*\rmdefault{bch}\normalfont\upshape
\rmfamily
\section*{}
\vspace{-1cm}

\begin{NoHyper}
\footnotetext{\textit{$^{*}$~Corresponding author.}}

\footnotetext{\textit{$^{a}$~Department of Physics, The Ohio State University, Columbus, OH 43210, USA. E-mail: bundschuh@mps.ohio-state.edu}}

\footnotetext{\textit{$^{b}$~Interdisciplinary Biophysics Graduate Program, The Ohio State University, Columbus, OH 43210, USA.}}
\footnotetext{\textit{$^{c}$~Department of Chemistry, Imperial College London, Molecular Sciences Research Hub, 80 Wood Lane, London W12 0BZ, UK.}}

\footnotetext{\textit{$^{d}$~Department of Mechanical and Aerospace Engineering, The Ohio State University, Columbus, OH 43210, USA.}}
\footnotetext{\textit{$^{e}$~Department of Chemistry and Biochemistry, The Ohio State University, Columbus, OH 43210, USA.}}
\footnotetext{\textit{$^{f}$~Division of Hematology, Department of Internal Medicine, The Ohio State University, Columbus, OH 43210, USA.}}
\footnotetext{\textit{$^{g}$~Center for RNA Biology, The Ohio State University, Columbus, OH 43210, USA.}}
\footnotetext{\dag~Electronic Supplementary Information (ESI) available. See DOI: 00.0000/00000000.}
\end{NoHyper}



\section{Introduction}

In 2006, Paul Rothemund published seminal work on the design of nanostructures out of DNA, developing a technique known as DNA origami \cite{rothemund2006folding}. Although early structures were static, expanding this technique to produce functional, dynamic structures has been of particular interest, since the use of DNA as a construction material renders the resulting structures naturally well-suited for use as machines in biological or synthetic systems. To this end, significant research has focused on the development of dynamic nanoscale devices~\cite{deluca2020dynamic, marras2015programmable}. Indeed, dynamic DNA origami devices are being developed for use as drug delivery systems~\cite{douglas2012logic, ijas2019reconfigurable, ketterer2016nanoscale}, as well as molecular biological probes~\cite{zhao2019quantitative, le2016probing}, computing elements, and nanorobots~\cite{deluca2020dynamic, gerling2015dynamic, kopperger2018self,nummelin2020robotic}. Interest in these applications has driven the development of a variety of actuation methods.  Actuation can be achieved in a number of ways, such as introduction of short oligonucleotides with specifically designed sequences~\cite{simmel2019principles, zhang2011dynamic, yurke2000dna}, or changing environmental factors such as salt conditions~\cite{marras2018cation, mao1999nanomechanical, gerling2015dynamic}, pH \cite{majikes2017ph, liu2006reversible,li2013motif, modi2009dna}, or temperature~\cite{arnott2019temperature, gerling2015dynamic, turek2018thermo}. 

In order to be suitable for use in such applications, however, it is necessary to have precise control over the stimulus response, which remains challenging. To this end, we quantitatively characterize the temperature actuation of a DNA origami hinge containing a gold nanoparticle (AuNP), which was previously described by Johnson \textit{et al.}~\cite{johnson2019reciprocal}. This device consists of two stiff arms connected by a flexible vertex, such that the motion around the vertex is restricted to a single angular dimension. A DNA-coated AuNP is attached to the top arm, and complementary DNA strands are affixed to the bottom arm that anneal to the AuNP to hold the hinge closed. When the temperature is increased, the hybridization between the AuNP and bottom arm melts to release the hinge into the open state. The AuNP remains stably attached to the top arm, allowing for repeatable temperature-controlled opening and closing. This type of system is of interest since AuNP-DNA origami composites have many exciting applications, such as in plasmonics~\cite{liu2018dna} and nanoelectronics~\cite{bayrak2018review}. In particular, the potential for AuNP facilitated reconfiguration that is both fast and tunable could be important in these applications. Another area of interest for these composite devices is that they have the potential to allow precisely controlled local heating and therefore actuation: Although the experiments described here are performed with bulk temperature change, the AuNP itself could in principle be locally heated with a laser~\cite{govorov2007generating}. Thus the study of systems into which such AuNPs are incorporated is a potentially fruitful area of research. 

In order to design increasingly complex and useful DNA devices, it is necessary to construct predictive models of their function. This has proven to be challenging, however, since environmental factors and thermal fluctuations can play an important role, often rendering classical solid mechanics approaches common to macroscopic engineering unsuitable~\cite{deluca2020dynamic}. Nevertheless, a number of computational techniques have shown predictive efficacy. All-atom molecular dynamics (MD) simulations, which track interactions of each atom in a system over time, provide detailed and accurate information about system dynamics~\cite{li2015ionic,wu2013molecular,yoo2013situ}. The computational cost is quite high, however, rendering such an approach practical only for small subsections of DNA devices and short time scales~\cite{deluca2020dynamic}. In order to study larger systems, one must use a coarse-grained approach. One way to do this is to approximate the atoms that make up a nucleotide as a single particle and track the positions and interactions of many such particles. This approach is taken most notably in the commonly used oxDNA simulation~\cite{ouldridge2011structural, vsulc2012sequence, snodin2015introducing, sharma2017characterizing}, as well as the recently developed MrDNA model~\cite{maffeo2020mrdna}. These approaches significantly extend simulation time-scales, but they are still typically limited to microsecond or at most millisecond timescales, while actuated conformational changes often occur on the second timescale or longer. Another coarse-grained approach is to use finite element (FE) modeling to predict DNA structures. A widely used example of this is CANDO~\cite{kim2012quantitative}, although this is typically used for shape prediction since it lacks molecular details that govern dynamics. Pan \textit{et al.} expanded the FE approach to improve the description of thermal fluctuations in shape, but this still did not extend to large-scale actuated conformational changes~\cite{pan2017structure}.

Long timescales or large structures may render even such coarse-grained approaches computationally unfeasible, particularly if one wants to rapidly iterate through many structural variations to guide design. It is therefore desirable to develop even more computationally efficient techniques. Here, we focus on the application of one such technique, statistical mechanics. While similar approaches have been used to model DNA strand displacement\cite{srinivas2013biophysics,irmisch2020modelling}, which is widely used for actuation, application to actuation of devices themselves remain rare despite the increasing need for computational efficiency to guide design of functional DNA origami devices. In one of the few examples of application to an actual device that we are aware of, Marras \textit{et al.} use a statistical mechanics approach to model a system in which changes in salt concentration are used to actuate a hinge~\cite{marras2018cation}. Here, we develop a thermodynamic model that accurately describes the temperature actuation of the nano-hinge device containing an AuNP. To our knowledge, this is the first statistical mechanics model of a \textit{composite} DNA origami system, which is a critical step, since many applications require the incorporation of NPs or other functional elements. Furthermore, we demonstrate that this model is able to predict actuation temperatures as a function of device design, enabling principled design of devices with desired transition temperatures.  Additionally, our model gives surprising insights into the system, revealing a limit on the number of simultaneously bound strands and demonstrating that configurational entropy and suboptimal energetic states meaningfully impact system behavior. 

\section{Experimental and computational methods}

\subsection{Experimental methods}

In this subsection we describe briefly the experiments by Johnson \textit{et al.} that provide the basis for our model~\cite{johnson2019reciprocal}. The AuNP-hinge system is shown schematically in Fig.~\ref{fgr:experiment}(A), with arrows indicating that the hinge is opened as temperature is increased and closed when temperature is decreased. In Fig.~\ref{fgr:experiment}(B), averaged data for different overhang strand lengths are shown for hinges with two overhang strands (left), which we call ``bivalent" or three overhang strands (right), which we call "trivalent". The overhang lengths (6-8 bases) and sequences (all adenine bases) are identical in the bivalent and trivalent cases. 

\subsubsection{Design and fabrication of DNA origami hinges}

The studied DNA origami hinges and AuNP-hinge constructs were prepared as previously described~\cite{johnson2019reciprocal}. Briefly, $20$ nM scaffold DNA (p8064) and $200$ nM staple strands were pooled in TE-buffer ($5$ mM Tris, $1$ mM EDTA, pH $8.0$, 5 mM NaCl) supplemented with $18$ mM MgCl$_2$ and subjected to a thermal annealing consisting of $15$ min at $65 \text{ }{}^{\circ}{}$C following by $4$ hours at $53 \text{ }{}^{\circ}{}$C and cooling to $4 \text{ }{}^{\circ}{}$C. Excess staple strands were removed by centrifugal purification in the presence of PEG~\cite{stahl2014facile}. Conjugation of T${}_{23}$ ssDNA coated AuNP, prepared as described by Johnson \textit{et al.}~\cite{johnson2019reciprocal}, to the purified DNA hinges was performed by addition of 5-fold excess AuNPs to the resuspended DNA hinges and incubation at $45\text{ }{}^{\circ}$C for 15 minutes.

\subsubsection{Thermal actuation}

Thermal profiles of the different constructs were collected on a Cary Eclipse Fluorometer with thermostated multicell cuvette holder. If not stated otherwise, temperature ramps were set to $2\text{ }{}^{\circ}$C/min and thermal profiles were collected by cycling between the respective minimum and maximum temperatures at least twice. A reference hinge without AuNP was used to substract temperature dependent fluorescence effects of the fluorophore.

\subsubsection{EM Imaging}

Negative stain electron microscopy was used to confirm folding and correct incorporation of AuNPs into the DNA hinges, following previously described protocols~\cite{johnson2019reciprocal}. Purified DNA hinges and AuNP-hinge constructs were adsorbed onto TEM grids (Electron Microscopy Sciences, Hatfield, PA), stained using freshly prepared Uranyl-formate and imaged on a FEI Tecnai G2 Spirit TEM, operated at $80$ kV.

\begin{figure}[ht]
\centering
  \includegraphics[width=\columnwidth]{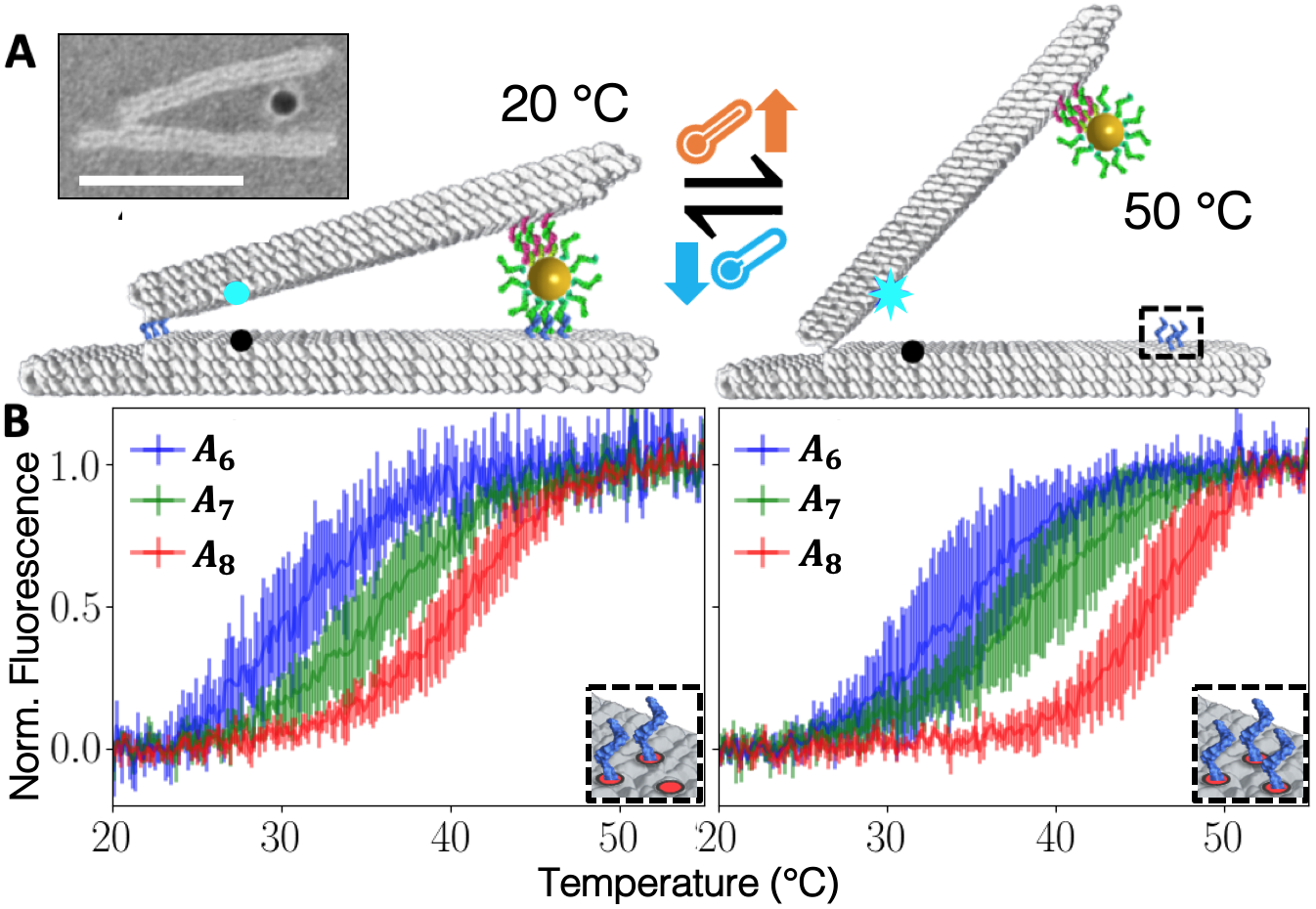}
  \caption{Experimental system and temperature actuation data underlying our model. (A) shows the experimental system: a gold nanoparticle (AuNP) is affixed to the top arm of a DNA origami nano-hinge via long, stable dsDNA strands formed between single stranded DNA (ssDNA) overhangs on the arm and complementary strands of ssDNA coating the AuNP.  Shorter overhangs on the bottom arm anneal at low temperatures and melt at high temperatures. The inset shows a TEM image of a closed hinge with AuNP at room temperature. Scale bar is $50$ nm. When the bottom overhang is annealed, the hinge is forced into a closed state where fluorescence is quenched. The normalized fluorescence in (B) therefore provides a measure of the bulk fraction of hinges that are open. The insets in (B) illustrate the number of overhangs on the bottom arm of the hinge: either two ("bivalent") or three ("trivalent"). The legend indicates the sequence of the bottom arm ssDNA, for instance $A_6$ corresponds to a sequence consisting of 6 adenine bases. All hinge overhangs are made up of adenine bases, and all AuNP connections are made up of thymine bases. The lengths of bottom arm overhangs vary between 6 and 8 in both the bivalent and trivalent cases.}
  \label{fgr:experiment}
\end{figure}

\subsection{Data processing}
\subsubsection{Normalization of fluorescence}
Following data collection and the substraction of fluorescence temperature dependence, the fluorescence is normalized such that the average maximum value (corresponding to all open hinges) is equal to one and the average minimum value (corresponding to all closed hinges) is equal to zero~\cite{johnson2019reciprocal}.  

\subsubsection{Averaging of Melting and Annealing Replicates}\label{sec:averaging}

The normalized data is averaged over all experimental replicates for both melting and annealing curves, and this average and the corresponding standard deviation are shown. Data from slightly different temperatures had to be combined due to fluctuations during thermal ramps which were set to collect one data point every $0.1\text{ }^\circ$C. Specifically, we use the temperature values of the first melting replicate and then identify the closest observed temperature values in other replicates to take the average and standard deviation of the corresponding fluorescence values. That average and standard deviation are assigned to the temperature value of the first melting replicate. In order to estimate the error due to the temperatures not lining up exactly among the replicates, we identified the maximum discrepancy in temperature values among replicates where fluorescence values are averaged, and compared the expected change in fluorescence according to our model to the experimental noise. We found that this worst case estimate of systematic error induced by averaging data from slightly different temperature values is on the order of the noise in the experimental measurements. Therefore, we concluded that any interpolation over temperature values from different replicates is unnecessary, as it would effectively constitute interpolation over noise. 

\subsection{Model calculation}

In this section, we describe how the experimental system is modeled, both at a conceptual and mathematical level. 

\subsubsection{Conceptual framework}

To relate the experimental readout to a calculable property of the system, we note that the fluorescence of nano-hinges, when normalized between $0$ and $1$, is equal to the fraction of open nano-hinges. Furthermore, for a system in thermodynamic equilibrium, the fraction of open hinges gives the probability that an \textit{individual} hinge is open. We therefore assume the system is in thermodynamic equilibrium and create a thermodynamic model of an individual hinge. 

\subsubsection{Model states and parameters}

\begin{figure}[htb]
\centering
  \includegraphics[width=\columnwidth]{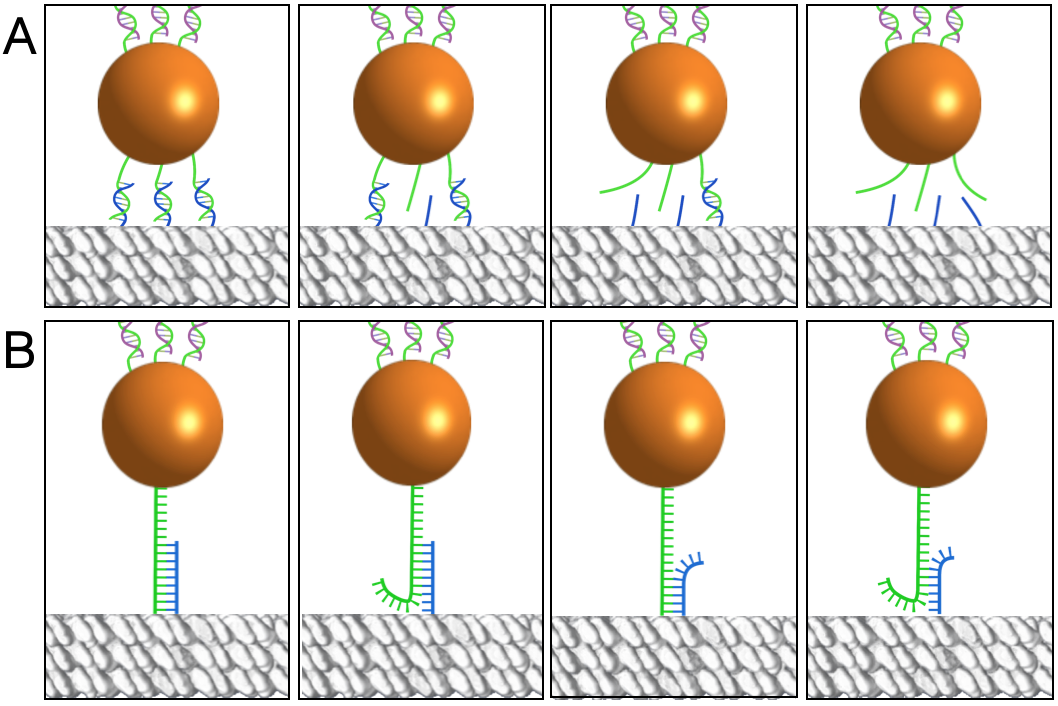}
  \caption{Schematic illustration of closed nano-hinge states. The AuNP is indicated by the gold sphere, the bottom hinge overhangs by blue lines, top arm overhangs by red lines, and the AuNP strands by green lines. The gray helices represent the bottom arm of the nano-hinge. (A) gives the states on the strand binding level, while (B) gives the states on the base pairing level. In (A) we note that there is one way to bind three connections (first panel), three ways to bind two connections (second panel), three ways to bind one connection (third panel), and one way to bind zero connections (fourth panel). This is described mathematically in Eq.~(\protect\ref{eq:Z_closed}). In (B), there are states with no fraying (first panel), states with fraying from the poly-T end (second panel), states with fraying from the poly-A end (third panel), and states with fraying from both ends (fourth panel). This is described mathematically in Eq.~(\protect\ref{eq:stack}).}
  \label{fgr:model}
\end{figure}

As discussed in more detail in section~\ref{sec:high_level_model}, a thermodynamic model requires enumeration of allowed states and corresponding free energies. The experimentally observable state in this system is whether the hinge is open or closed, so we consider the microstates of the system that correspond to these macrostates. We treat an open hinge as consisting of only a single state, capturing the effect of the many physical microstates in the closing free energy parameters, the hinge closing enthalpy change $\Delta H_{cl}$ and the hinge closing entropy change $\Delta S_{cl}$. For a closed hinge, we enumerate the possible binding states more explicitly as shown in Fig.~\ref{fgr:model}. First, any number of the bottom overhangs on the nano-hinge can be involved in base pairing with the ssDNA strands on the AuNP (Fig~\ref{fgr:model}A).  For any given base-pairing interaction, any consecutive stretch of adenines on the bottom overhang can bind to any consecutive stretch of equal length of thymines on the ssDNA strands on the AuNP  (Fig~\ref{fgr:model}B).  Each set of consecutive base pairs is associated with a base pairing enthalphy $\Delta H_{bp}$ and entropy $\Delta S_{bp}$. 

We do not consider any base-pairing states that involve bulges or internal loops, i.e., unpaired bases internal to a base-paired region, which have a prohibitively high free energy cost. Various experimental results indicate that the free energy cost of such states is at least on the order of $4$ kJ/mol. Tanaka \textit{et al.}~\cite{tanaka2004thermodynamic} find that free energies for single A and T bulges at $37^{\circ}$C are $6.95 \pm 4.39$ kJ/mol and $4.0 \pm 3.9$ kJ/mol, respectively. Longer bulges are expected to be similarly costly, as reported by Turner and Matthews in the context of RNA~\cite{turner2010nndb}. The cost of interior loops can be approximated using mismatch parameters for an ACA/TTT sequence, which Allawi \textit{et al.} find to be $5.82 \pm 0.46$ kJ/mol~\cite{allawi1998thermodynamics}. Similarly, Peyret \textit{et al.} find AA and TT mismatches to have energy costs on the order $4$ kJ/mol~\cite{peyret1999nearest}. At a cost of at least $4$ kJ/mol, the Boltzmann factor corresponding to a 7 base polyA-polyT section of dsDNA drops from $\approx 700$ to below $10$ upon the creation of a bulge or interior loop. 

\subsubsection{Opening probability}\label{sec:openprob}

In order to compute the probability that the system resides in the states specified above, it is necessary to determine the free energies associated with each state and to compute a partition function. The free energy of each of these microstates is determined by the number of paired bases and the energy required to close the hinge. Since the temperature is variable, both the enthalpic and entropic parts of the free energies must be considered. Using the open state as a reference free energy $G_\textrm{open} = 0$, we can write
\begin{equation}
    G_i(T) = \Delta H_{cl} - T\Delta S_{cl} + N_i(\Delta H_{bp} - T \Delta S_{bp}).
    \label{eq:G_enthalpy_and_entropy}
\end{equation}
for each closed microstate $i$, where T is temperature, $N_i$ is the number of base stacks (since the base pairing energy is associated with the energetic favorability of stacking two consecutive base pairs); $\Delta H_{cl}$
and $\Delta S_{cl}$ are the hinge closing enthalpy and entropy, respectively; and $\Delta H_{bp}$
and $\Delta S_{bp}$ are the base pairing  enthalpy and entropy, respectively. The partition function for this system is then
\begin{equation}
    Z = 1+\sum_i \exp{\big[-G_i(T) / k_B T\big]},
\end{equation} where the sum is over all microstates of the system. 
Since we treat the hinge as consisting of a single open state with free energy $G_\textrm{open} \equiv 0$, the probability that a hinge is in an open state is given by
\begin{equation}
p_{\mathrm{open}} = 1 / Z.
\label{eq:p_open}
\end{equation}

In order to calculate $Z$ more explicitly, we have to consider the partition function of the closed states, denoted $Z_{\textrm{cs}}$. Since all closed states are multiplied by a Boltzmann factor corresponding to the free energy cost to close the hinge, denoted by $\Delta G_{cl} = \Delta H_{cl} - T \Delta S_{cl}$, the primary challenge is to account for all possible base pairing states, as illustrated in  Fig.~\ref{fgr:model}. To do this for an arbitrary number of overhangs per hinge with potentially differing numbers of bases per overhang, we first denote overhangs of different lengths (i.e. different numbers of bases) by subscript $j$, and the number of type $j$ overhangs by $N_{\mathrm{c},j}$. For instance, if we have a hinge with three overhangs, two of which with six bases (type 1) and one of which with eight bases (type 2), we would have $N_{\mathrm{c},1}=2$ and $N_{\mathrm{c},2}=1$. We then need to account for all possible choices of actually realized connections $n_j$ of type $j$ subject to the constraint $\sum_j n_j \leq N_{\mathrm{c,}\, \mathrm{max}}$, where $N_{\mathrm{c,}\, \mathrm{max}}$ is the maximum number of bound connections per state. The number of possibilities to choose $n_j$ out of $N_{\mathrm{c,}j}$ total available connections must also be considered, introducing a binomial coefficient for each $j$. Combining the above yields 
\begin{multline}
Z_{\textrm{cs}} = \exp{\big( - \Delta G_{cl} / k_B T \big)} \times \\ \sum_{n_1=0}^{\min\{N_{\mathrm{c,}\, \mathrm{max}},\, N_{\mathrm{c,}1}\}} \Bigg[\binom{N_{\mathrm{c,}1}}{n_1} Z_{\mathrm{S,}1}^{n_1}  \times\!\!\!\!\!\!\!\!\!\!\!\! \sum_{n_2=0}^{\min\{N_{\mathrm{c,}\, \mathrm{max}} - n_1,\, N_{\mathrm{c,}2}\}} \Bigg[\binom{N_{\mathrm{c,}2}}{n_2} Z_{\mathrm{S,}2}^{n_2} \times \cdots  \Bigg] \Bigg]
\label{eq:Z_closed}
\end{multline} where $Z_{\mathrm{S,}j}$ is the partition function describing all of the possible  base pairing interactions for a single connection of type $j$. In particular,  
\begin{equation}
Z_{\mathrm{S,}j} \equiv \sum_ {i = 1}^{N_{\mathrm{S,}j}} (N_T - i + 1) (N_{\mathrm{A,}j} - i +1) \exp\left[-\frac{\Delta G_\textrm{term} + i \Delta G_{bp}}{k_B T} \right]
\label{eq:stack}
\end{equation} where $N_{\mathrm{S,}j}$ is the maximum number of stacks in the type $j$ duplex, $N_T$ is the maximum number of stacks available to the poly-T strand, and $N_{\mathrm{A,}j}$ is the maximum number of stacks available to the poly-A strand of type $j$. Thus, $N_{\mathrm{S,}j} = \min\{ N_T, N_{\mathrm{A,}j} \}$. $\Delta G_{bp} = \Delta H_{bp} - T \Delta S_{bp}$ is the free energy of a single stack, and $\Delta G_{term} = 2\big[9.6 \textrm{ kJ/mol} - T(0.0172 \textrm{ kJ /(mol K)} )\big]$ is the terminal base pairing energy~\cite{santalucia1998unified}. Note that $N_T - i + 1$ is the number of positions on the poly-T strand at which \textit{i} consecutive stacks can bind, and that $N_{\mathrm{A,}j} - i +1$ is the number of positions on the poly-A strand at which \textit{i} consecutive stacks can bind. Thus, their product is the total multiplicity of the state with \textit{i} bound stacks.

Taking everything together, we therefore have 
 \begin{multline}
 Z =  1+ \exp{\big( - \Delta G_{cl} / k_B T \big)} \times \\ \sum_{n_1=0}^{\min\{N_{\mathrm{c,}\, \mathrm{max}},\, N_{\mathrm{c,}1}\}} \Bigg[\binom{N_{\mathrm{c,}1}}{n_1} Z_{\mathrm{S,}1}^{n_1}  \times\!\!\!\!\!\!\!\!\!\!\!\! \sum_{n_2=0}^{\min\{N_{\mathrm{c,}\, \mathrm{max}} - n_1,\, N_{\mathrm{c,}2}\}} \Bigg[\binom{N_{\mathrm{c,}2}}{n_2} Z_{\mathrm{S,}2}^{n_2} \times \cdots  \Bigg] \Bigg]
 \label{eq:Z_detailed}
  \end{multline} with $p_{\mathrm{open}} = 1/Z$, where for clarity the $\Delta G$'s are not written as functions of temperature, but they retain the temperature dependence indicated in Eq.~(\ref{eq:G_enthalpy_and_entropy}).

\subsection{Model fitting}\label{sec:model_fitting}

In order to relate the model to the data, we fit the opening probability, $p_{\mathrm{open}}$, to the experimental normalized fluorescence by varying the four energetic parameters within physically realistic bounds (i.e. $0\leq \Delta H_{cl} \leq \infty$ and $-\infty \leq \Delta S_{cl}, \Delta H_{bp}, \Delta S_{bp} \leq  0$). This fit is performed via a non-linear least squares minimization for this bounded set of parameters using a Trust Region Reflective algorithm~\cite{branch1999subspace}, which is implemented using the Python SciPy package~\cite{2020SciPy-NMeth}. Additionally, we performed a discrete optimization over the maximal number of overhangs $N_{\mathrm{c,max}}\in\{1,2,3\}$ available for simultaneous binding. 

\subsection{Code and data availability}

Python code implementing the model is available at \url{https://github.com/bundschuhlab/PublicationScripts/tree/master/NanoswitchTActuationPrediction}. Data is available upon request. 

\section{Results \& Discussion}

In this section, we will first give a short high level overview of our model and then demonstrate how it agrees with the experimental data and expectations based on the literature.  Next, we extract mechanistic insights about nano-hinge actuation. Lastly, we demonstrate that the model can be used to guide the design of nano-hinges that can be actuated over a wide range of temperatures. 

\subsection{A thermodynamic model of thermal nano-hinge actuation}\label{sec:high_level_model} 

We formulate a thermodynamic model for actuation of the hinge containing an AuNP as shown in Fig.~\ref{fgr:experiment}(A).  Specification of a thermodynamic model requires enumeration of the allowed states of the system, the free energies associated with each state, as well as a relationship between these states and experimental observables. The macroscopic observable here is the open (fluorescing) or closed (quenched) state of the hinge. We model the system as having a single open state, representing many microstates, which has some unknown free energy cost to transition into the closed state. This free energy cost captures the contribution from all open microstates and consists of an enthalpic component $\Delta H_{cl}$ and an entropic component $\Delta S_{cl}$. When the hinge is closed, there are many binding microstates available, but these states are easier to enumerate: when closed, the hinge overhangs are allowed to anneal to the AuNP DNAs, and every combination of consecutive base pairing stacks is allowed. This is shown schematically in Fig.~\ref{fgr:model}.  Fig.~\ref{fgr:model}(A) shows the strand-level allowed binding states, and Fig.~\ref{fgr:model}(B) shows the base pair-level allowed binding states. Fraying is allowed from the AuNP strand end, the hinge overhang end, and from both ends. Additionally, the strands are allowed to slide relative to each other without penalty, so that any combination of consecutive bases can anneal (all the way up to the AuNP). Since the hinge overhangs are poly-A and the AuNP overhangs are poly-T, each stack of two consecutive base pairs that forms is associated with the same base pairing free energy with enthalpic component $\Delta H_{bp}$ and entropic component $\Delta S_{bp}$.  Given these definitions of the states and their free energies, the partition function of the system and thus the probability of a hinge to be in the open state and fluorescing can be calculated as a function of temperature (see section~\ref{sec:openprob}).

\subsection{Nano-hinge actuation is quantitatively explained by the thermodynamic model}\label{sec:firstfit}

We fit the temperature dependent opening probability predicted by the thermodynamic model to the experimental normalized fluorescence using a non-linear least squares minimization for the enthalpies and entropies $\Delta H_{cl}$, $\Delta S_{cl}$, $\Delta H_{bp}$, and $\Delta S_{bp}$ and discrete optimization over the maximal number of overhangs $N_{\mathrm{c,max}}\in\{1,2,3\}$ available for simultaneous binding (see section~\ref{sec:model_fitting}). To test the predictive power and robustness of the model we first fit to bivalent (two overhangs on the bottom arm of the hinge) and trivalent (three overhangs on the bottom arm of the hinge) data simultaneously using all five fit parameters. We then fit to bivalent and trivalent data separately varying the four energy parameters but keeping constant the value of $N_{\mathrm{c,max}}=2$ found to be optimal in the simultaneous fit. 

\begin{figure}[htb]
\centering
  \includegraphics[width=\columnwidth]{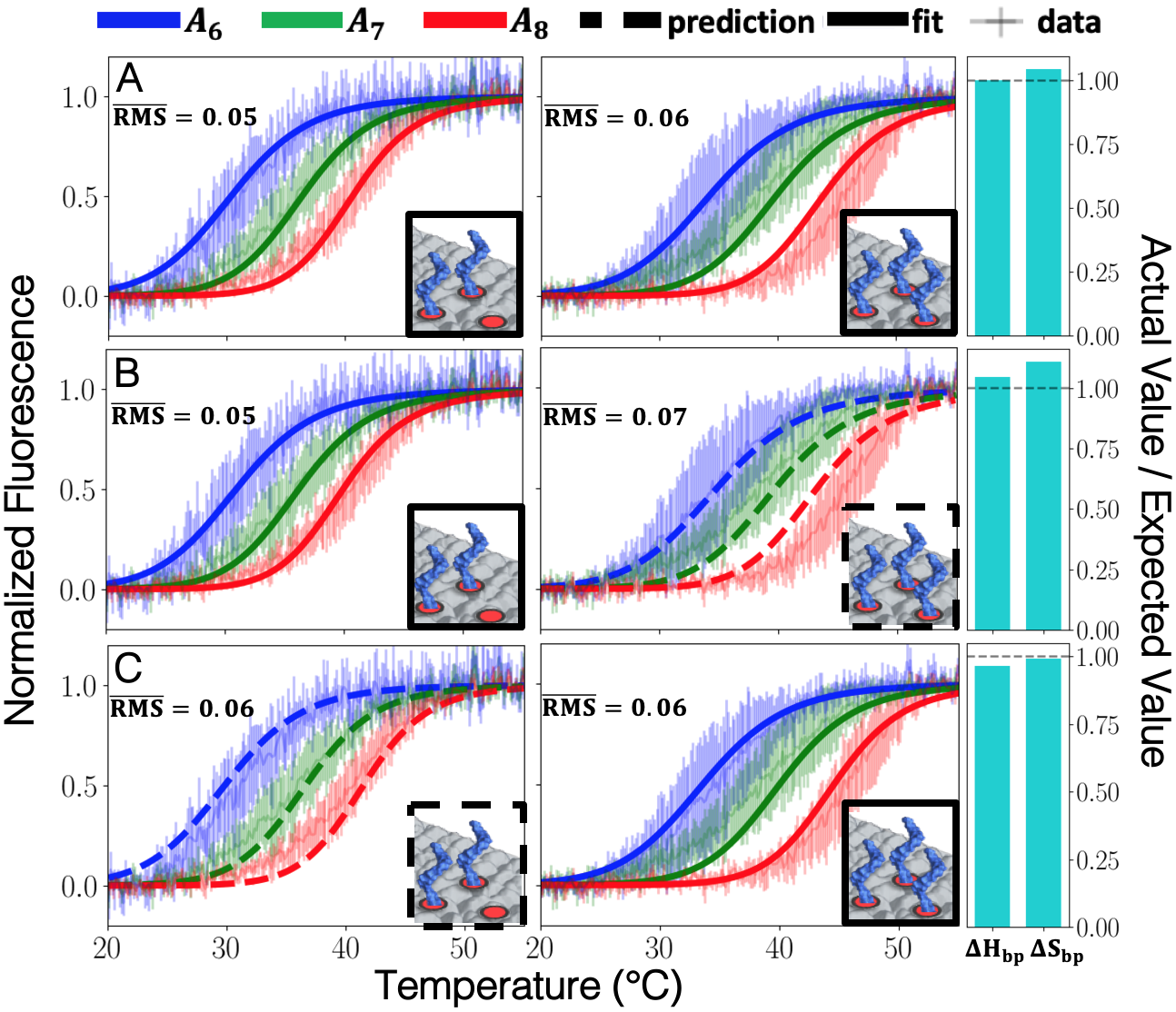}
  \caption{Model with $N_{\mathrm{c,max}}=2$ fit to experimental data. In row (A), the model is fit to bivalent and trivalent data simultaneously. In row (B), the model is fit to the bivalent data (solid lines), and the trivalent model with bivalent fit parameters is compared to trivalent data (dashed lines). In row (C), the model is fit to the trivalent data (solid lines), and the bivalent model with trivalent fit parameters is compared to bivalent data (dashed lines). Each panel contains the average root mean squared difference ($\overline{\textrm{RMS}}$) between the model and the experimental data. The rightmost column shows the ratio between the fit base pairing parameter values and the expected base pairing parameter values~\cite{santalucia1998unified}. }
  \label{fgr:results}
\end{figure}

A comparison of these fits to the data is shown in Fig.~\ref{fgr:results}, and the best fit parameters are summarized in Table~\ref{tbl:fit_vals}. In Fig.~\ref{fgr:results}, row (A) shows the simultaneous fit to the bivalent and trivalent data, row (B) shows the fit to the bivalent data and the predictions for the trivalent system using the bivalent fit parameters, and row (C) shows the fit to the trivalent data and the predictions for the bivalent system using the trivalent fit parameters. Each panel shows the average root mean squared difference ($\overline{\textrm{RMS}}$) between the data and model. In all cases, the model agrees well with the data with a maximum $\overline{\textrm{RMS}}$ of $0.06$ for the fit data and of $0.07$ for the predicted curves.

\begin{table}[ht]
\footnotesize
  \caption{Best fit parameter values: Model parameters used for curves shown in Fig.~\ref{fgr:results} in units of kJ/mol for enthalpies and kJ/mol K for entropies.}
  \label{tbl:fit_vals}
  \begin{tabular*}{\columnwidth}{@{\extracolsep{\fill}}llllll}
    \hline
    Data fit & $\Delta H_{cl}$ & $\Delta S_{cl}$ & $\Delta H_{bp}$ & $\Delta S_{bp}$ & $N_{c,\mathrm{max}}$  \\
    \hline
    Tri. \& Bi. & $0 \pm 17$ & $-0.097 \pm -0.056$ & $-33.1 \pm 2.1$ & $-0.100 \pm 0.007$ & 2  \\
    Bi. & $0 \pm 19$ & $-0.087 \pm 0.062$ & $-34.6 \pm 2.4$ & $-0.106 \pm 0.008$ & 2 \\
    Tri. & $0 \pm 32$ & $-0.108 \pm 0.102$ & $-31.9 \pm 3.7$ & $-0.095 \pm 0.012$ & 2 \\
    \hline
  \end{tabular*}
\end{table}

We want to emphasize that, while the experimental data is separated by valency rather than overhang length, the experimental actuation curves for the bivalent and trivalent cases differ (as shown in Fig.~\ref{fgr:tri_and_bi_data_comp}) and thus provide independent tests for the model.  The fits in Fig.~\ref{fgr:results}(B-C) show that only fitting to either the bivalent or trivalent data allows prediction of the other with nearly the same quality as fitting to both. Thus, the valency based difference is accurately captured by the model without further adjustment of its parameters. We conclude that the thermodynamic model faithfully describes the entire temperature dependence of nano-hinge actuation for six different experimental conditions spanning two different valencies and three different overhang lengths using five fit parameters.

\subsection{Optimal base pairing parameters agree with literature values}

While we treat the enthalpy $\Delta H_{bp}$ and entropy $\Delta S_{bp}$ of the base pairing as fit parameters, these have been independently measured by SantaLucia~\cite{santalucia1998unified}  from melting experiments on short DNA oligomers and have been used for decades to quantitatively describe DNA melting. Therefore, it is illustrative to  compare our best fit parameters to SantaLucia's values. The rightmost column of Fig.~\ref{fgr:results} shows that the ratio of our best fit parameters to SantaLucia's values with corrections (details in ESI section~\ref{sec:salt_correction}\dag) for salt concentration~\cite{owczarzy2008predicting} for each of the three fits is close to one. As a further test of the appropriateness of the observed  base pairing parameters, we fit the model again while keeping the base pairing enthalpy $\Delta H_{bp}$ and entropy $\Delta S_{bp}$ at SantaLucia's literature values, corrected for salt conditions~\cite{owczarzy2008predicting,record1978semiempirical, santalucia1998unified,erie1987dumbbell}  (details in ESI section~\ref{sec:salt_correction}\dag). These fits are shown in Supplementary Fig.~\ref{fgr:fix_stacking_params}, and the fit parameters are given in Table~\ref{tbl:fix_stacking_params}. We find an excellent fit if only the base pairing enthalpy is fixed to its literature value and a reasonable fit if base pairing enthalpy and entropy are both fixed at their literature values. These fits are especially reasonable when considering that the effects of divalent salt remain difficult to quantify and mostly affect the entropy~\cite{record1978semiempirical, santalucia1998unified, erie1987dumbbell}.

The good agreement between literature values of the  base pairing parameters and the best fit parameters of our model is interesting, since other studies have found that the presence of a DNA origami device can have significant impact on  base pairing free energy~\cite{marras2018cation,shi2020free}.  One possible explanation is that in the experiments underlying these earlier studies the base pairing occurs in a much more geometrically constrained context, which is avoided by the presence of the AuNP coated with longer DNA strands in the experiments that are modeled here. Also, the corrections in these studies are directly to the base pairing free energy~\cite{marras2018cation,shi2020free} while we consider enthalphy and entropy separately to model the entire temperature dependence.  Since the free energy results from a delicate balance between enthalpic and entropic contributions, the free energy may be more sensitive than enthalpy or entropy alone.  We conclude that our fit parameters and literature values for the base pairing parameters agree well, providing further independent validation of our model.

\subsection{Averaging over melting and annealing data approximates equilibrium conditions}\label{sec:hysteresis}

For our fitting we use the data published by Johnson \textit{et al.}~\cite{johnson2019reciprocal} averaged over both experimental replicates and direction of temperature change, with the width of the curves corresponding to the standard deviation over all of these data sets (see section~\ref{sec:averaging} for details).  Although we model this system as an equilibrium process, it is important to note that there \textit{is} hysteresis in the experimental data between the annealing, which exhibits a slightly lower transition temperature, and the melting, which exhibits a slightly higher transition temperature. In order to verify that the average of these melting and annealing curves is a good approximation to an equilibrium condition, we perform the thermal actuation experiment at two different rates of temperature change: $2\text{ }{}^{\circ}$C/min (as previously done by Johnson \textit{et al.}~\cite{johnson2019reciprocal}) and $0.2\text{ }{}^{\circ}$C/min. For these experiments, we replaced the AuNP with double-stranded DNA linkers to avoid potential AuNP degradation with extended time at elevated temperatures~\cite{li2013thermal}. These experiments reveal that as the rate is decreased, the hysteresis also decreases and both the melting and annealing curves approach the average of the fast rate curves. Additionally, the average of the slow rate curves is similar to the average of the fast rate curves. This data, shown in Fig.~\ref{fgr:comp_rate_data}, illustrates that the averaging is a reasonable approximation to the equilibrium conditions.

Nevertheless, we also test the model in the two extreme assumptions that the true equilibrium is either the melting or annealing data. These fits are shown in Fig.~\ref{fgr:melt_and_anneal_only}, with best fit parameters reported in Table~\ref{tbl:melt_and_anneal_only}. While these fits result in $\overline{\textrm{RMS}}$ values that are somewhat higher than the fits to averaged data (0.07 and 0.08 for bivalent and trivalent, respectively), the values of the base pairing parameters still agree well with literature values, particularly when relying on just the annealing. We conclude that our observations concerning the validity of the thermodynamic model are robust to the details of the treatment of the experimentally observed hysteresis.

\subsection{At most two overhangs participate simultaneously in base pairing}

The parameter $N_{\mathrm{c,max}}$ of our model is the maximal number of overhangs that participate in base pairing simultaneously. In the bivalent case, only two overhangs are present, but since in the trivalent case three overhangs are present, it is somewhat surprising that $N_{\mathrm{c,max}} = 2$ provides the best fit to our model.  In order to evaluate if $N_{\mathrm{c,max}} = 2$ really provides a significantly better fit of the experimental data than the more natural $N_{\mathrm{c,max}} = 3$, we fit the model again varying the free energy fit parameters $\Delta H_{cl}$, $\Delta S_{cl}$, $\Delta H_{bp}$, and $\Delta S_{bp}$ but keeping the maximal number $N_{\mathrm{c,max}} = 3$ of simultaneous overhangs fixed. These fits are shown in Supplementary Fig.~\ref{fgr:Ncmax3_fits}, with parameter values given in Table~\ref{tbl:fit_vals_Ncmax3}.  The simultaneous fit to all six experimental conditions is of low quality ($\overline{\textrm{RMS}}=0.1$ in both the bivalent and trivalent cases). Fitting the bivalent and trivalent data alone gives good fits with $\overline{\textrm{RMS}}$ values of 0.05 and 0.06, respectively (note that $N_{\mathrm{c,max}} = 3$ is identical to $N_{\mathrm{c,max}} = 2$ for the bivalent case since only two overhangs can bind). However, when only one of these data sets is fit, the prediction of the data set excluded from the fit is completely inconsistent with the experimental results.  In addition, the best fit values for the base pairing enthalpy and entropy are very inconsistent with SantaLucia's literature values except for the case of the fit to the bivalent data alone, where $N_{\mathrm{c,max}}=2$ and $N_{\mathrm{c,max}}=3$ are equivalent. We therefore conclude that even in the presence of three overhangs, simultaneous binding must be constrained to at most two of these overhangs.

For completeness, we also show the fits for a maximal number $N_{\mathrm{c,max}}=1$ of simultaneously binding overhangs in Supplementary Fig.~\ref{fgr:Ncmax1_fits}, with parameter values in Table~\ref{tbl:fit_vals_Ncmax1}. These fits and/or the agreement with literature expectation of their resulting base pairing parameter values are poor, so we conclude that more than one overhang must be involved in simultaneous binding.

Although a determination of the mechanism of the constraint of at most two simultaneously bound overhangs is beyond the scope of this study, there are a number of possible explanations. Previously, Shi \textit{et al.} observed a confinement effect via molecular simulation that is in general favorable to base pairing~\cite{shi2020free}. However, Marras \textit{et al.} observed a significant decrease in  base pairing favorability due to strand confinement upon comparison of a statistical mechanical model to experiment~\cite{marras2018cation}, and Jonchhe \textit{et al.} observed a large decrease in stability of duplex DNA when a DNA hairpin is confined to a 15x15 nm nanocage~\cite{jonchhe2020duplex}. They attributed this reduced stability to reduced water activity that results from the strong alignment of water molecules with the charged environment~\cite{jonchhe2018decreased, jonchhe2020duplex}. These studies suggest that localization of the overhang DNA strands due to hybridization could introduce a confinement effect that decreases the favorability of base pairing, 
which in turn could set an effective strand limit on binding. In our case, the decrease in base pairing favorability could be driven by physical distortions of the preferred dsDNA geometry in a confined space (i.e. the the dsDNA may be forced into unfavorable bends or twists in order to bind properly). Additionally, there may be an increasingly high energy cost to bring more negative charges (both the DNA bases and AuNP are negatively charged) into a small space, in particular if the screening is reduced by confinement, as suggested by Jonchhe \textit{et al.} There also may be single stranded AuNP strands within the space enclosed by the three hinge overhangs, the hinge arm, and the bottom surface of the nanoparticle. If this were the case, the configurational entropy of these strands could decrease sharply upon binding of a third strand, rendering this energetically unfavorable. 

It is also possible that there are simply only two AuNP DNA strands available to bind to the overhangs, or that the the number of available AuNP strands in the sample have some distribution such that they on average behave as if two strands are available. Such an effect could originate from a low density of DNA on the AuNP. Finally, the geometry of the connections on top of the hinge may fix the position of the AuNP relative to the bottom hinge arm in such a way that only (effectively) two strands are able to access the bottom arm overhangs. 

\subsection{Difference between bivalent and trivalent cases is driven by configurational entropy}

Having a validated thermodynamic model allows elucidation of the mechanisms of actuation, including the difference between the bivalent and the trivalent case. Since the maximum number of simultaneously bound overhangs is $N_{\mathrm{c,max}}=2$, it may be surprising that there is a difference between the bivalent and trivalent cases at all, since the lowest free energy state is the same in {\em both} cases, corresponding to two overhangs being fully base paired with complementary strands on the AuNP.  Yet, the experimental data clearly differs between the bivalent and the trivalent case, and the model quantitatively captures this difference. The model reveals that the difference between the bivalent and the trivalent case comes solely from configurational entropy: in the bivalent system there are two possible ways one overhang can bind the AuNP and one way two overhangs can bind the AuNP; in the trivalent system there are three possible ways one overhang can bind the AuNP and three possible ways two overhangs can bind the AuNP. Thus, there are more microstates in the trivalent system than in the bivalent system, even if the energetics of these states are identical. In fact, if all three overhangs were able to bind, a much larger change in actuation temperatures would be expected (Fig.~\ref{fgr:Ncmax3_fits}). Therefore, we conclude that the additional configurational entropy in the trivalent case causes the shifts of the actuation points toward higher temperatures.

\subsection{Fraying plays a measurable role in nano-hinge actuation}

The thermodynamic model also elucidates the role of sliding and fraying in the binding between hinge overhangs and strands attached to the AuNP. As shown in Fig.~\ref{fgr:model}(B) the model takes into account fraying of the base pairing between hinge and AuNP overhangs at each end as well as arbitrary sliding of the two strands relative to each other (since they are homopolymers).  Supplementary Fig.~\ref{fgr:no_slide_no_fray} (parameters in Table~\ref{tbl:fit_vals_no_slide_no_fray}) shows fits for variants of the model in which fraying and sliding (A), fraying (B), and sliding (C) are not allowed (see ESI section~\ref{sec:nofraying}\dag).  While the $\overline{\textrm{RMS}}$ values are similar to those in Fig.~\ref{fgr:results}, the agreement with SantaLucia's base pairing parameters is significantly worse for the cases in which fraying is not allowed, suggesting suboptimal annealing states are important in regulating the thermal actuation. 

Interestingly, however, the sliding states do not seem to play as important a role. The $\overline{\textrm{RMS}}$ values are again similar to those in Fig.~\ref{fgr:results}, but here the entropic base pairing parameter is closer to expectation while the enthalpic one shows a bigger discrepancy. We are more confident about the expected value of the enthalpic base pairing parameter, since it should not be impacted by salt~\cite{record1978semiempirical, santalucia1998unified, erie1987dumbbell}; hence, we think it likely that sliding is having some impact that the model is capturing. Since the extent to which sliding influences the actuation is unknown, we conclude that it is prudent to leave sliding states in the model. 

\subsection{Quantitative model allows design of devices with arbitrary transition temperatures}

\begin{figure}[htb]
\centering
  \includegraphics[width=0.6\columnwidth]{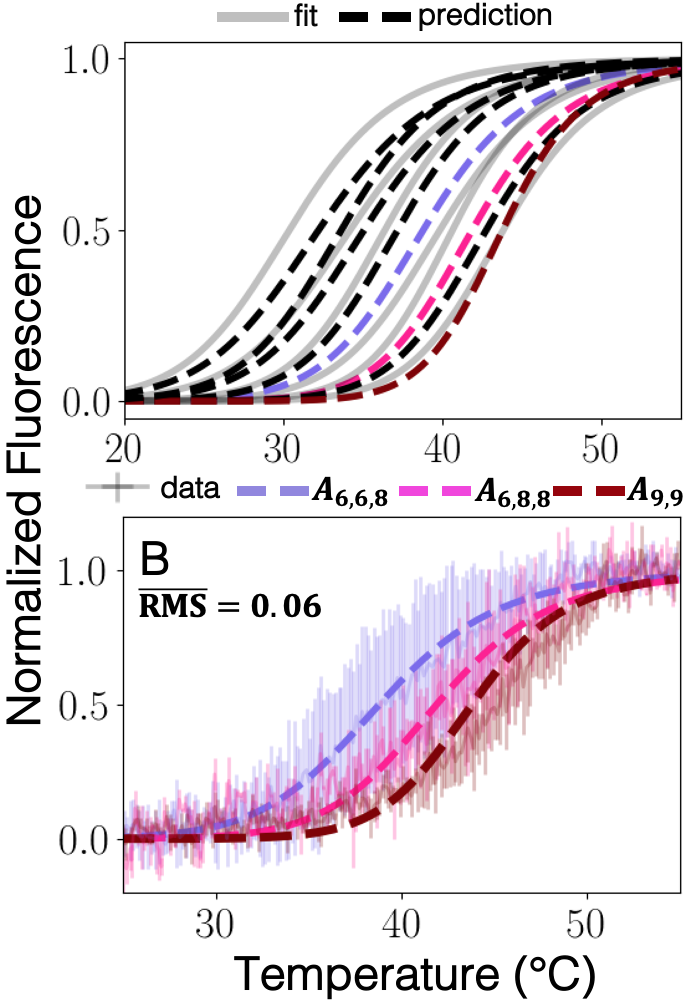}
  \caption{(A) Temperature spread of predicted actuation curves for hypothetical nano-hinges with variable overhang length and valency. A complete description of the designs sorted by actuation temperature can be found in Table~\protect\ref{tbl:melt_temps}. Solid and grayed lines are the curves fit to data (trivalent and bivalent 6, 7, and 8 base overhangs), while dashed lines indicate predictions. The violet, pink, and maroon dashed lines indicate predictions validated (in the same colors) in (B). As mentioned above, (B) gives validation of the $A_{6,6,8}$, $A_{6,8,8}$, and $A_{9,9}$ predictions via a comparison to experimental data (plotted with thin lines in the same colors) which were not used to create or fit the model.}
  \label{fgr:discussion}
\end{figure}

The model shows excellent agreement with the data, and is able to predict the change in actuation response due to a change in overhang valency and length. This predictive power may be useful to achieve actuation at a desired temperature, avoiding costly experimental trial and error. the model predicts that manipulation of overhang design parameters can be used to achieve actuation at essentially any desired temperature in the range from about $30\text{ }{}^{\circ}\mathrm{C}$ to $45\text{ }{}^{\circ}\mathrm{C}$ as shown Fig.~\ref{fgr:discussion}(A). The predicted actuation temperatures of each design are included in Table~\ref{tbl:melt_temps}.

To validate the ability to guide design, we show in Fig.~\ref{fgr:discussion}(B) that three of these predicted actuation curves agree very well ($\overline{\textrm{RMS}}=0.06$) with experimental data that were not used in model development. These data correspond to AuNP nano-hinges with: two 9-base polyA overhangs ($A_{9,9}$), previously published by Johnson \textit{et al.}~\cite{johnson2019reciprocal} ; two 6-base overhangs and one 8-base overhang ($A_{6,6,8}$); and two 8-base overhangs and one 6-base overhang ($A_{6,8,8}$). The two latter data sets are original to this work and thus their raw melting and annealing curves are shown in Fig.~\ref{fgr:mixed_melt_and_anneal_data}.  We conclude that the model can be used to design overhang combinations with essentially arbitrary actuation temperatures in the range from about $30\text{ }{}^{\circ}\mathrm{C}$ to $45\text{ }{}^{\circ}\mathrm{C}$.

\begin{table}[ht]
\small
  \caption{Actuation temperatures (i.e. the temperatures at which the model predicts that the hinge is equally likely to be open or closed) of hinges shown in Fig.~\protect\ref{fgr:discussion}. Each number corresponds to the length of an overhang, so designs with two numbers are bivalent hinges and designs with three numbers are trivalent hinges. Actuation temperatures can be tuned to within $2\text{ }{}^{\circ}$C of any desired temperature between $30\text{ }{}^{\circ}$C and $45\text{ }{}^{\circ}$C}
  \label{tbl:melt_temps}
  \begin{tabular*}{\columnwidth}{@{\extracolsep{\fill}}lclc}
    \hline
    Design & $T_{\textrm{act}}$ ($^{\circ}$C) &Design & $T_{\textrm{act}}$ ($^{\circ}$C)    \\
    \hline
    6,6 & 30.4  & 6,6,8 & 38.8  \\
     5,6,6  & 32.2  & 7,7,7 & 39.9   \\
    6,7 & 33.7  & 8,8 & 40.6   \\
     6,6,6  & 34.2  & 6,8,8 & 41.9    \\
    5,6,7 & 35.1  & 7,8,8  & 42.8  \\
   7,7 & 36.4 &  8,8,8 & 43.8 \\
   4,5,9 & 37.2 &  9,9 & 43.8  \\
    \hline
  \end{tabular*}
\end{table}

We also considered the possibility of extrapolation of the model to higher and lower temperatures. Johnson \textit{et al.}, however, did not observe transitions in the case of a trivalent hinge with 9-base overhangs ($A_{9,9,9}$) up to $55\text{ }{}^{\circ}\mathrm{C}$~\cite{johnson2019reciprocal}. This disagrees with the model prediction of an equilibrium actuation temperature of $46.8\text{ }{}^{\circ}\mathrm{C}$ for $A_{9,9,9}$,  so it would seem that the model should not be extrapolated to temperatures above the previously indicated $45\text{ }{}^{\circ}\mathrm{C}$. 

Since the nano-hinges themselves melt around $60\text{ }{}^{\circ}\mathrm{C}$~\cite{wei2013mapping, castro2011primer, johnson2019reciprocal}, it is not entirely unexpected that the model begins to break down at higher temperatures. As the temperature increases and individual sections of the hinge begin to melt, the hinge may become more flexible, effectively decreasing $\Delta G_{cl}$ and thus increasing the probability of closed states. Additionally, some local melting of the connections that hold the nano-hinge overhangs in place could grant them more flexibility and remove the $N_\textrm{c,max} = 2$ constraint. If this constraint were removed with all other best fit parameters kept the same, the model would predict $A_{9,9,9}$ actuation at $57.3\text{ }{}^{\circ}\mathrm{C}$, which is consistent with the experimental data.

At lower temperatures, Johnson \textit{et al.} did not observe actuation in either a bivalent or trivalent hinge with 5-base overhangs ($A_{5,5}$ and $A_{5,5,5}$) down to $20\text{ }{}^{\circ}\mathrm{C}$~\cite{johnson2019reciprocal}. This represents a discrepancy of only a few degrees with the model prediction of equilibrium actuation temperatures of $21.4\text{ }{}^{\circ}\mathrm{C}$ and $25.7\text{ }{}^{\circ}\mathrm{C}$ for $A_{5,5}$ and $A_{5,5,5}$, respectively. Since it is unlikely that the nano-hinge changes significantly at the lower end of the temperature range, this discrepancy could be due to slower kinetics at lower temperatures, which may result in hinges never fully closing on experimental timescales.

\section{Conclusions}

The fast, accurate, and predictive thermodynamic DNA origami actuation model developed in this work offers a viable alternative to computationally costly molecular dynamics modeling in the design of dynamic DNA origami devices. We have shown that not only is it useful as a design tool, but it is able to provide mechanistic insight into the actuation process that suggest future avenues of experimental research.  In particular, determination of the precise physical origin of the $N_{\textrm{c,max}}=2$ constraint may increase understanding of composite DNA nano-structures. For example, experiments that examine the effect of variation of density of the DNA coating the AuNP would be interesting, although this is challenging to control precisely in practice. Furthermore, the creation of increasingly complex  dynamic DNA devices necessitates increasingly computationally efficient modelling~\cite{deluca2020dynamic}, of which statistical mechanics is likely to be an important part. This type of model is in principle applicable to any device that is actuated by melting/annealing of DNA duplexes, and similar methods have been shown to be applicable to other dynamic structures~\cite{marras2018cation}. Attempts to use statistical mechanics methods to model a wider range of devices, as well as the incorporation of kinetics using transition matrices acting on microstates (such as the ones defined here) to capture non-equilibrium effects, are important areas of future research.

\section*{Conflicts of interest}
There are no conflicts to declare.

\section*{Acknowledgements}
We thank the Winter lab for providing nanoparticles for the additional experiments and valuable feedback on the work. This material is based upon work supported by the National Science Foundation under Grant No.~DMR-1719316 to RB and by the Department of Energy under Grant no.~DE-SC0017270 to CC.

\balance

\bibliography{np_paper} 
\bibliographystyle{rsc} 

\cleardoublepage

\section{Supplementary material}

\nobalance

\renewcommand\thefigure{S\arabic{figure}}    
\setcounter{figure}{0}

\renewcommand\thetable{S\arabic{table}}    
\setcounter{table}{0}

\renewcommand\textfraction{0}
\setcounter{totalnumber}{10}

\subsection{Salt corrected base pairing parameters}\label{sec:salt_correction}

We calculate the salt corrected base pairing parameter values using the melting temperature correction
\begin{multline}
    \frac{1}{T_m(\textrm{Mg}^{2+}\!)} = \frac{1}{T_m(\textrm{1 M Na}^+\!)}\! +\! a\! +\! b \ln [\textrm{Mg}^{2+}] \!+\! f_{GC}(c\!+\!d\ln[\textrm{Mg}^{2+}]) + \\ \frac{1}{2 (N_{bp} - 1)}\bigg\{e+f\ln[\textrm{Mg}^{2+}]+g(\ln[\textrm{Mg}^{2+}])^2\bigg\}
\end{multline}
given by Owczarzy {et al.}~\cite{owczarzy2008predicting}, where $a$, $b$, $c$, $d$, $e$, $f$, and $g$ are experimentally determined constants. We use the magnesium salt correction since our and Johnson \textit{et al.}'s experiments are performed in the presence of $11.5$ mM free magnesium. Since we care about the change in  base pairing energy for a generic internal base, we ignore edge effects by taking the limit as $N_{bp}$ approaches infinity. Furthermore, our sequences consist solely of A's and T's, so the fraction of G and C bases $f_{GC}=0$. Thus the melting temperature in our case is given by
\begin{equation}
    \frac{1}{T_m(\textrm{Mg}^{2+})} = \frac{1}{T_m(\textrm{1 M Na}^+)} + \big(3.92 - 0.911 \ln [0.0115]\big)\times 10^{-5}\label{eq:Tm}
\end{equation} where we have substituted the appropriate constants for $a$ and $b$. 

Since the melting temperature is the temperature at which the  base pairing energy change is 0, we can write the melting temperature in terms of the  base pairing entropy and enthalpy change as 
\begin{equation}
    T_m = \Delta H_{bp} / \Delta S_{bp}.
\end{equation}
Assuming that salt concentration only impacts  base pairing entropy~\cite{record1978semiempirical, santalucia1998unified, erie1987dumbbell}, the corrected base pairing entropy change $\Delta S_{bp,c}$ is given by 
\begin{equation}
    \Delta S_{bp,c} = \frac{\Delta H_{bp}}{T_m(\textrm{Mg}^{2+})}\label{eq:salt_correction}
\end{equation}
with $T_m(\textrm{Mg}^{2+})$ from Eq.~(\ref{eq:Tm}).

\subsection{Modeling the effect of fraying or sliding}\label{sec:nofraying}

In order to disentangle the effects of fraying and sliding of the DNA duplexes between the overhangs and the DNA strands on the AuNP, we use alternative models that exclude each of them separately or both of them. Fraying and sliding enter the original model via the multiplicities in Eq.~(\ref{eq:stack}). We thus here provide alternative versions of Eq.~(\ref{eq:stack}) for the three different cases that exclude fraying and/or sliding.

\subsubsection{Model with no fraying or sliding}

In order to exclude fraying and sliding, Eq.~(\ref{eq:stack}) becomes
\begin{equation}
Z_{\mathrm{S,}j} =  \exp\left[ -\frac{\Delta G_\textrm{term} + N_{\mathrm{S,}j} \Delta G_s}{k_B T} \right], 
\end{equation} since there is now only a single state allowed with free energy $\Delta G_\textrm{term} + N_{\mathrm{S,}j} \Delta G_s$.

\subsubsection{Model with no fraying}
In order to exclude only fraying, Eq.~(\ref{eq:stack}) becomes
\begin{equation}
Z_{\mathrm{S,}j} =  (|N_{\mathrm{T}} - N_{\mathrm{A,}j}|+1) \exp\left[-\frac{\Delta G_\textrm{term} + N_{\mathrm{A,}j} \Delta G_s}{k_B T} \right], 
\end{equation} where the prefactor is simply the number of positions available to completely paired strands.   

\subsubsection{Model with no sliding}
In order to exclude only sliding, Eq.~(\ref{eq:stack}) becomes
\begin{equation}
Z_{\mathrm{S},j} =  \sum_ {i = 1}^{N_{\mathrm{S,}j}}  (N_{\mathrm{S,}j}-i+1)\exp\left[-\frac{\Delta G_\textrm{term} + i \Delta G_s}{k_B T} \right],
\end{equation} where the sum is over the number of possible base paired stacks, and the prefactor gives the multiplicity of each of these stacks. 

\subsection{Supplemental tables}

\begin{table}[h]
\footnotesize
  \caption{Fixed base pairing parameters in units of kJ/mol for enthalpies and kJ/mol K for entropies. Values without errors are fixed.}
  \label{tbl:fix_stacking_params}
  \begin{tabular*}{\columnwidth}{@{\extracolsep{\fill}}llllll}
    \hline
    Data fit & $\Delta H_{cl}$ & $\Delta S_{cl}$ & $\Delta H_{bp}$ & $\Delta S_{bp}$ & $N_{c,\mathrm{max}}$  \\
    \hline
    Enthalpy fixed & $0 \pm 1.7$ & $-0.097 \pm 0.006$ & $ -33.1$ & $-0.0998 \pm 0.0001$ & 2  \\
    Both fixed & $0 \pm 2.6$ & $-0.137 \pm 0.008$ & $-33.1$ & $-0.0955$ & 2  \\
    \hline
  \end{tabular*}
\end{table}

\begin{table}[h]
\footnotesize
  \caption{Melt and anneal only parameters in units of kJ/mol for enthalpies and kJ/mol K for entropies.}
  \label{tbl:melt_and_anneal_only}
  \begin{tabular*}{\columnwidth}{@{\extracolsep{\fill}}llllll}
    \hline
    Data fit & $\Delta H_{cl}$ & $\Delta S_{cl}$ & $\Delta H_{bp}$ & $\Delta S_{bp}$ & $N_{c,\mathrm{max}}$  \\
    \hline
    Melt & $0 \pm 27$ & $-0.099 \pm 0.087$ & $ -36.0 \pm 3.2$ & $-0.108 \pm 0.010$ & 2  \\
    Anneal & $0 \pm 27$ & $-0.108 \pm 0.088$ & $ -33.1 \pm 3.1$ & $-0.099 \pm 0.010$ & 2  \\
    \hline
  \end{tabular*}
\end{table}

\begin{table}[h]
\footnotesize
  \caption{$N_{\mathrm{c,max}} = 3$ fit parameters in units of kJ/mol for enthalpies and kJ/mol K for entropies.}
  \label{tbl:fit_vals_Ncmax3}
  \begin{tabular*}{\columnwidth}{@{\extracolsep{\fill}}llllll}
    \hline
    Data fit & $\Delta H_{cl}$ & $\Delta S_{cl}$ & $\Delta H_{bp}$ & $\Delta S_{bp}$ & $N_{c,\mathrm{max}}$  \\
    \hline
    Tri. \& Bi. & $0 \pm 11$ & $-0.036 \pm 0.035$ & $-75.9 \pm 3.5$ & $-0.250 \pm 0.011$ & 3  \\
    Bi. & $0 \pm 19$ & $-0.087 \pm 0.062$ & $-34.6 \pm 2.4$ & $-0.106 \pm 0.008$ & 3 \\
    Tri. & $0 \pm 30$ & $-0.113 \pm 0.095$ & $-27.3 \pm 2.7$ & $-0.083 \pm 0.009$ & 3 \\
    \hline
  \end{tabular*}
\end{table}

\begin{table}[h]
\footnotesize
  \caption{$N_{\mathrm{c,max}} = 1$ fit parameters in units of kJ/mol for enthalpies and kJ/mol K for entropies.}
  \label{tbl:fit_vals_Ncmax1}
  \begin{tabular*}{\columnwidth}{@{\extracolsep{\fill}}llllll}
    \hline
    Data fit & $\Delta H_{cl}$ & $\Delta S_{cl}$ & $\Delta H_{bp}$ & $\Delta S_{bp}$ & $N_{c,\mathrm{max}}$  \\
    \hline
    Tri. \& Bi. & $0 \pm 22$ & $-0.082 \pm 0.070$ & $-47.2 \pm 4.3$ & $-0.139 \pm 0.014$ & 1  \\
    Bi. & $0 \pm 22$ & $-0.079 \pm 0.072$ & $-51.1 \pm 4.5$ & $-0.153 \pm 0.014$ & 1 \\
    Tri. & $0 \pm 27$ & $-0.094 \pm 0.085$ & $-48.9 \pm 5.1$ & $-0.142 \pm 0.016$ & 1 \\
    \hline
  \end{tabular*}
\end{table}

\begin{table}[h]
\scriptsize
  \caption{No fraying and no sliding fit parameters in units of kJ/mol for enthalpies and kJ/mol K for entropies.}
  \label{tbl:fit_vals_no_slide_no_fray}
  \begin{tabular*}{\columnwidth}{@{\extracolsep{\fill}}llllll}
    \hline
    Data fit & $\Delta H_{cl}$ & $\Delta S_{cl}$ & $\Delta H_{bp}$ & $\Delta S_{bp}$ & $N_{c,\mathrm{max}}$  \\
    \hline
    No slide or fray & $0 \pm 14$ & $-0.040 \pm 0.044$ & $ -26.9 \pm 1.4$ & $-0.079 \pm 0.004$ & 2  \\
    No fray & $0 \pm 15$ & $-0.079 \pm 0.050$ & $ -24.3 \pm 1.4$ & $-0.071 \pm 0.004$ & 2 \\
    No slide & $0 \pm 16$ & $-0.056 \pm 0.050$ & $-31.7 \pm 1.8$ & $-0.095 \pm 0.006$ & 2 \\
    \hline
  \end{tabular*}
\end{table}

\newpage

\subsection{Supplemental figures}

\begin{figure}[h]
\centering
\includegraphics[width=0.6\columnwidth]{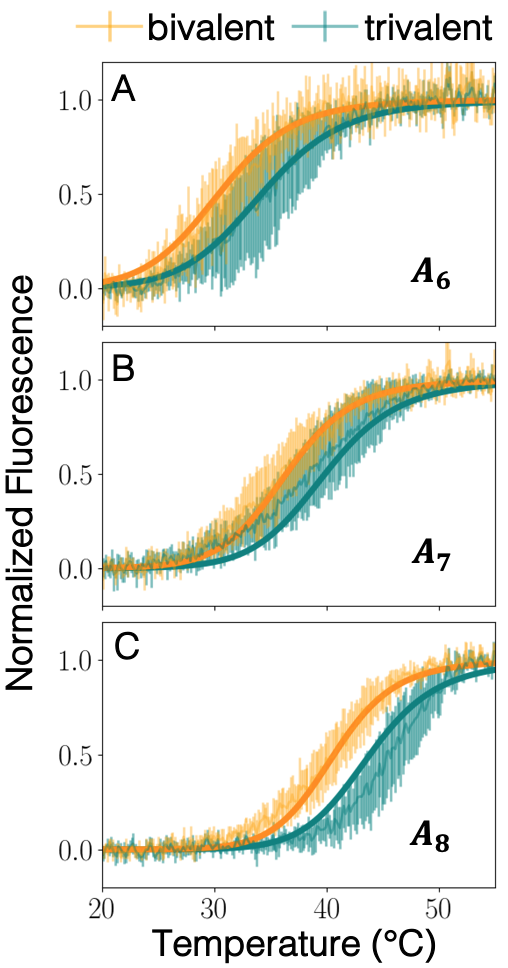}
\caption{Effect of overhang strand valency on actuation curves. The experimental data (thin lines) and model fits (thick lines) are identical to the ones shown in Fig.~\protect\ref{fgr:results}(A) but rather than separating the data by valency, each panel shows the bivalent (orange) and the trivalent (turquoise) case for (A) 6 base polyA overhangs, (B) 7 base polyA overhangs, and (C) 8 base polyA overhangs.}
\label{fgr:tri_and_bi_data_comp}
\end{figure}

\begin{figure}[h]
\centering
  \includegraphics[width=\columnwidth]{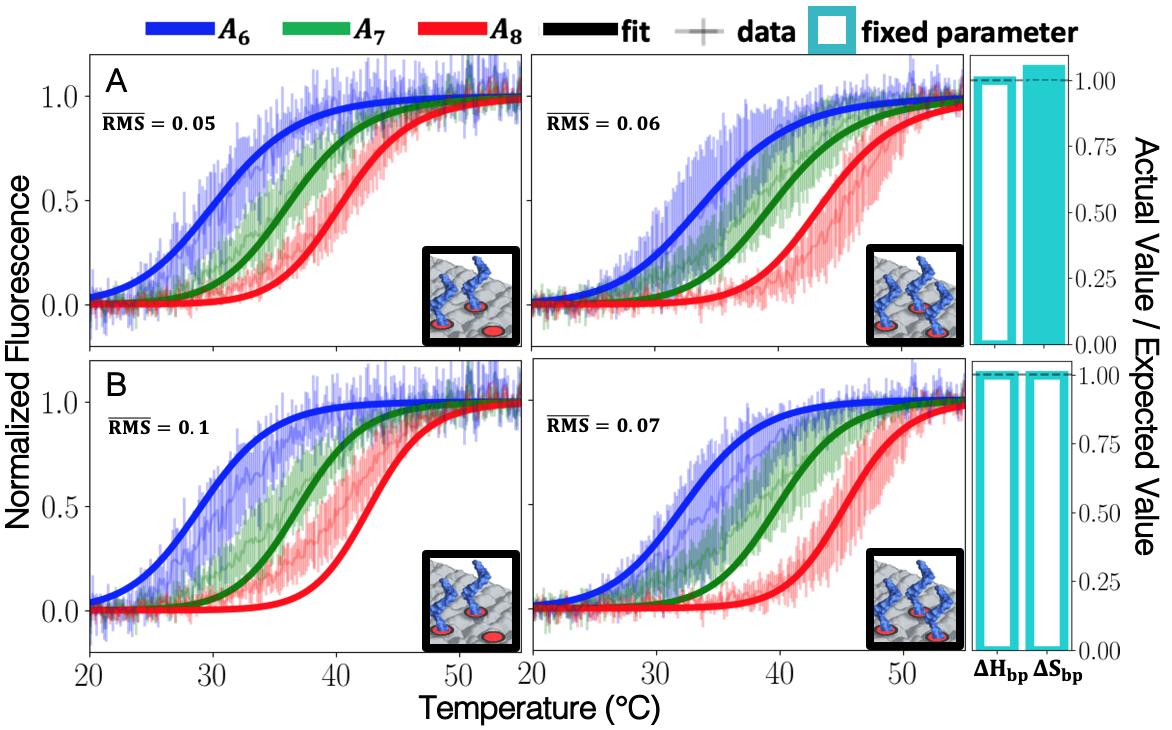}
  \caption{Model with $N_{\mathrm{c,max}}=2$ fit to experimental datasets. In (A),  base pairing enthalpy $\Delta H_{bp}$ is fixed at the SantaLucia expected value, and in (B) both the  base pairing enthalpy and the salt corrected  base pairing entropy $\Delta S_{bp}$ are fixed. In both cases, the model is fit to bivalent and trivalent data simultaneously. Each panel contains the average root mean squared difference ($\overline{\textrm{RMS}}$) between the model and the average of the experimental data. The rightmost column shows the ratio between the fit base pairing parameter values and the salt corrected expected base pairing parameter values given by Eq.~\ref{eq:salt_correction}. Since the outlined bars are fixed to the expected values, they have an actual to expected value ratio of 1.}
  \label{fgr:fix_stacking_params}
\end{figure}

\begin{figure}[h]
\centering
  \includegraphics[width=\columnwidth]{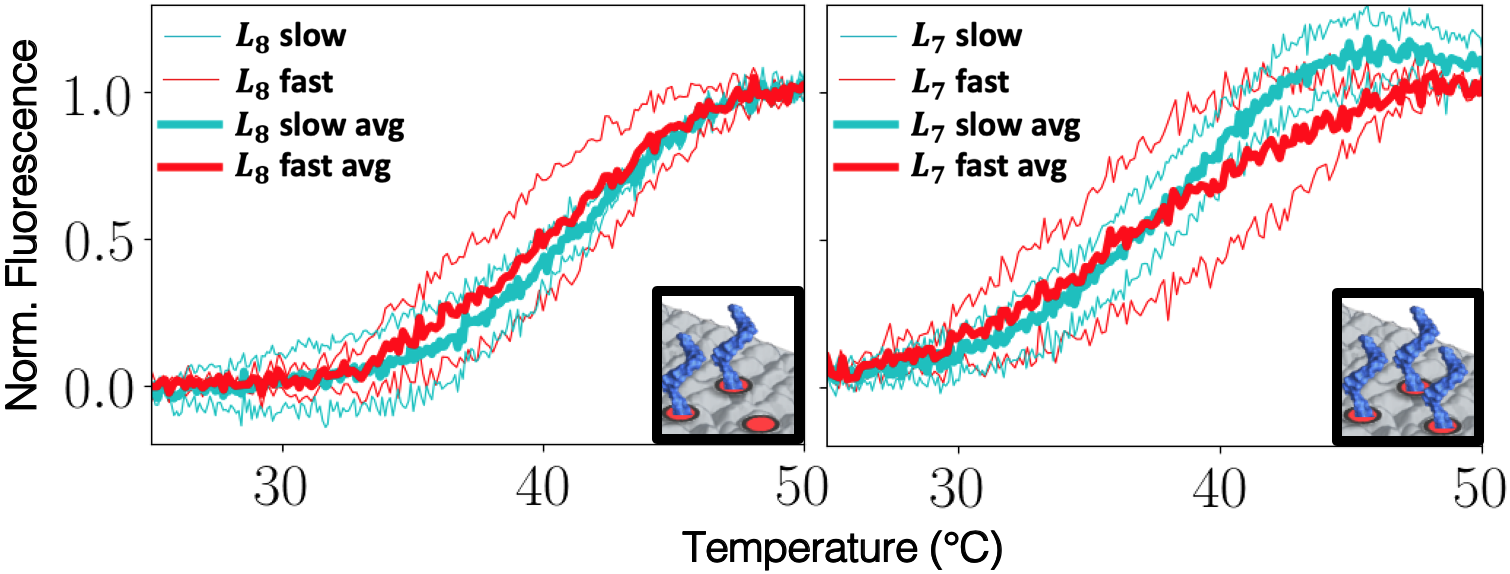}
  \caption{Comparison of melting and annealing curves for linker DNA (with no AuNP) with varied rates. These experiments are similar to those with DNA nano-hinge containing AuNP, except that the duplexes from the top arm are extended and the AuNP removed. These duplexes have polyT single-stranded portions on the ends that mimic the single stranded polyTs that normally coat the AuNP. In particular, fabrication included hinges with 10 fold excess of linkers ($200$ nM) using the same folding protocol as for all other hinges. The left panel shows melting and annealing curves for two linkers annealing to 8 base overhangs, while the right panel shows melting and annealing curves for three linkers annealing to 7 base overhangs. For the red curves the temperature is changed at a rate of $2^\circ \textrm{C} / \textrm{min}$, while for the green curves the temperature is changed at a rate of $0.2^\circ \textrm{C} / \textrm{min}$.}
  \label{fgr:comp_rate_data}
  
\end{figure}

\begin{figure}[h]
\centering
  \includegraphics[width=\columnwidth]{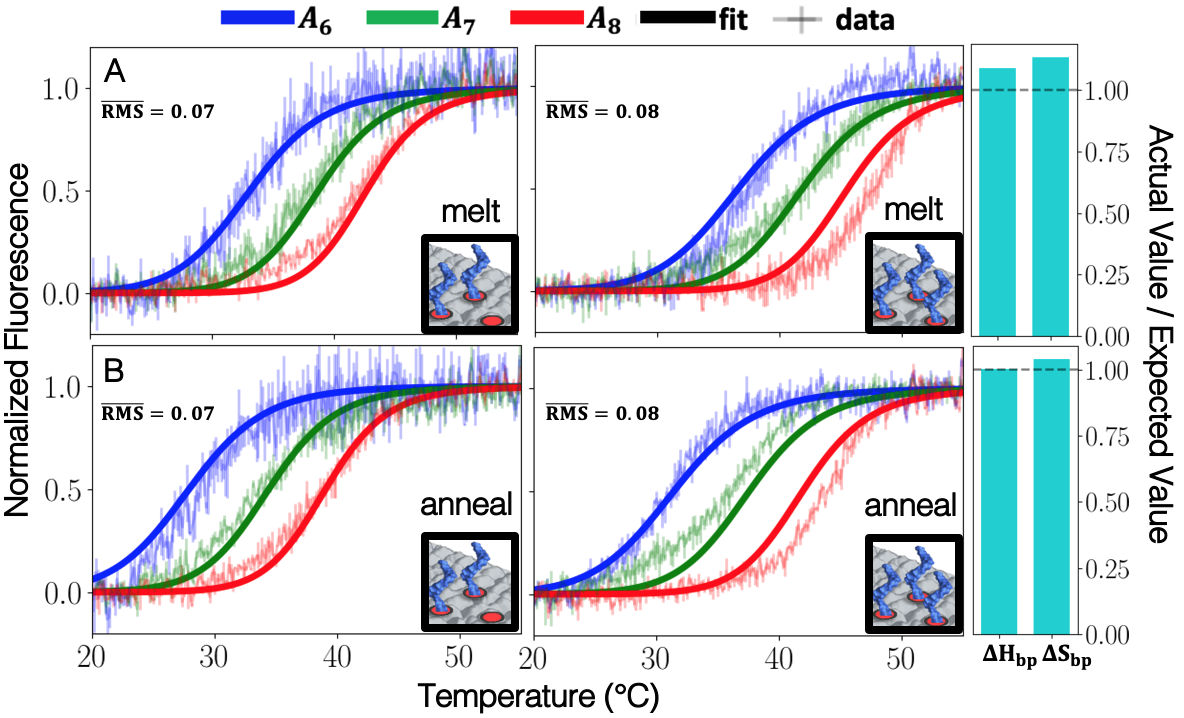}
  \caption{Model with $N_{\mathrm{c,max}}=2$ fit to experimental datasets containing only melting (A) and only annealing (B) data. In both cases, the model is fit to bivalent and trivalent data simultaneously. Each panel contains the root mean squared difference ($\overline{\textrm{RMS}}$) between the model and the average of the experimental data. The rightmost column shows the ratio between the fit base pairing parameter values and the expected base pairing parameter values~\cite{santalucia1998unified}.}
  \label{fgr:melt_and_anneal_only}
\end{figure}

\begin{figure}[h]
\centering
  \includegraphics[width=\columnwidth]{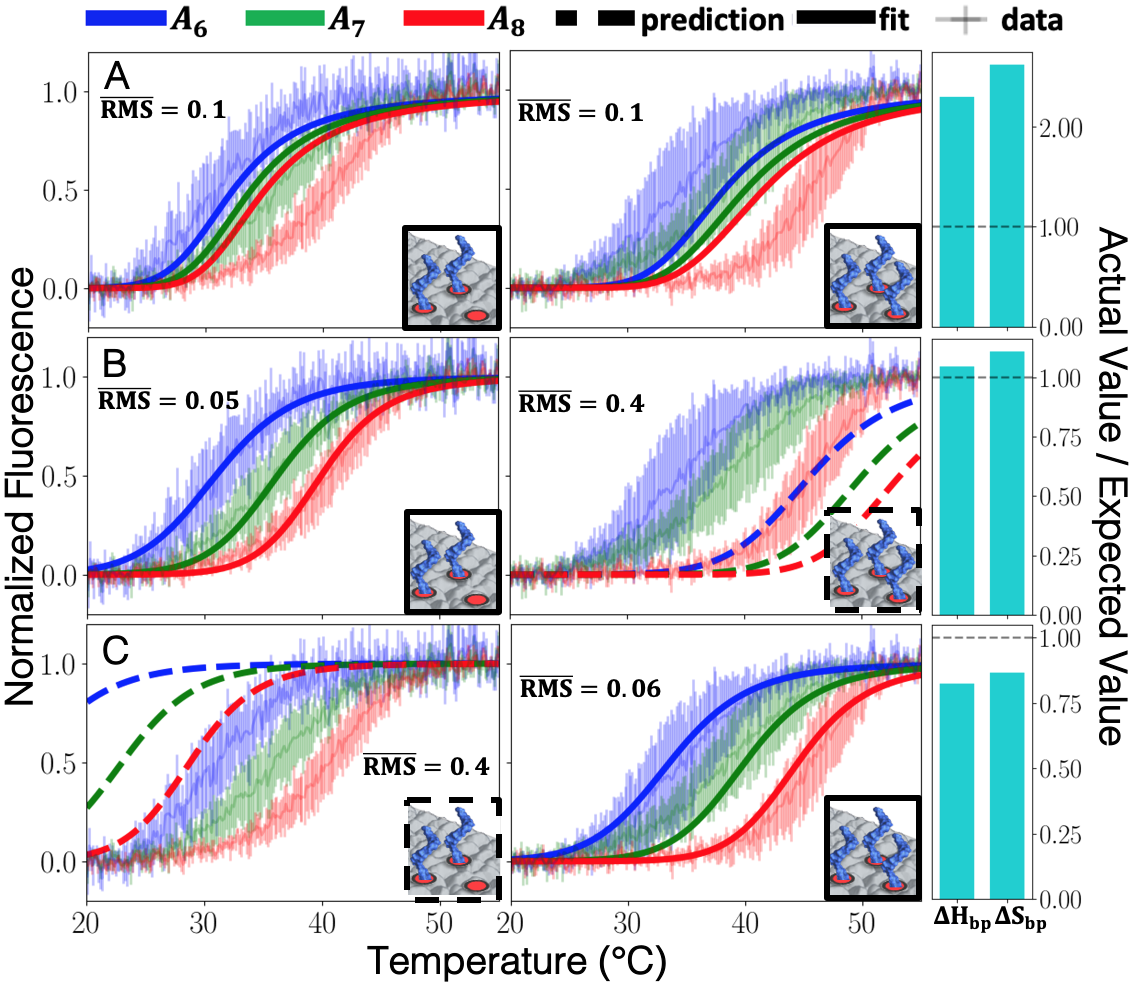}
  \caption{Model with $N_{\mathrm{c,max}}=3$ fit to experimental data. In row (A), the model is fit to bivalent and trivalent data simultaneously. In row (B), the model is fit to the bivalent data (solid lines), and the trivalent model with bivalent fit parameters is compared to trivalent data (dashed lines). In row (C), the model is fit to the trivalent data (solid lines), and the bivalent model with trivalent fit parameters is compared to bivalent data (dashed lines). Each panel contains the average root mean squared difference ($\overline{\textrm{RMS}}$) between the model and the average of the experimental data. The rightmost column shows the ratio between the fit base pairing parameter values and the expected base pairing parameter values~\cite{santalucia1998unified}.}
  \label{fgr:Ncmax3_fits}
\end{figure}

\begin{figure}[h]
\centering
  \includegraphics[width=\columnwidth]{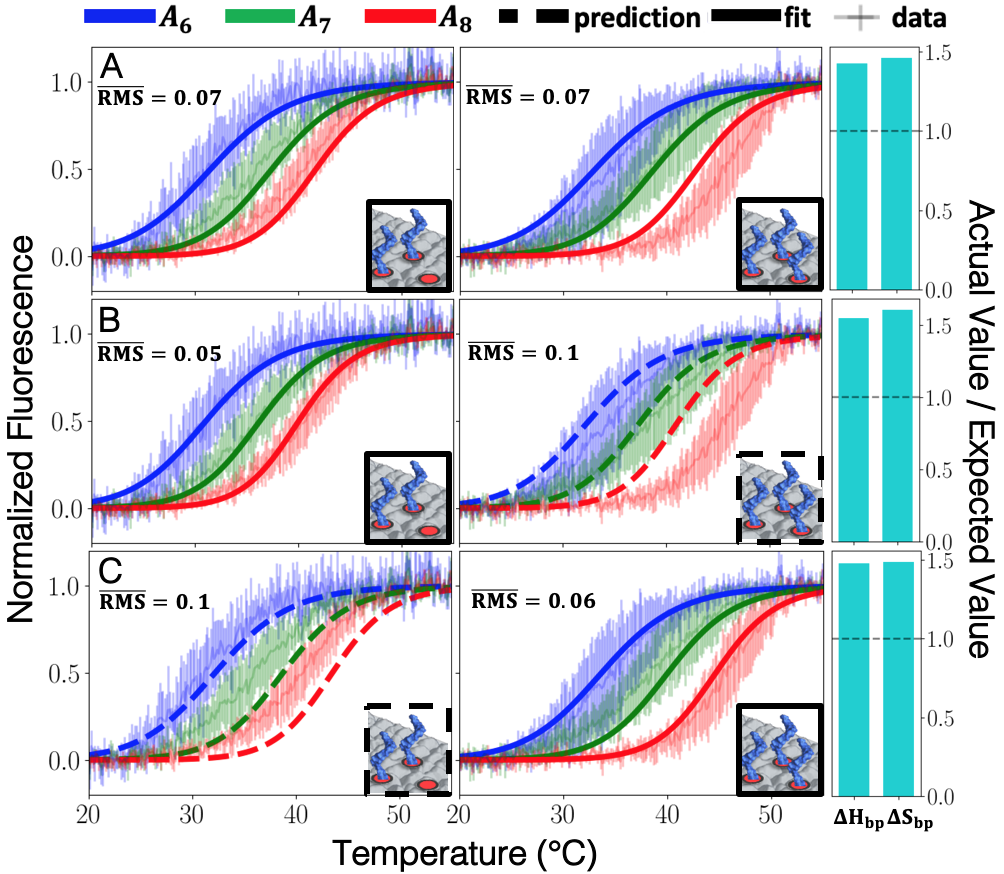}
  \caption{Model with $N_{\mathrm{c,max}}=1$ fit to experimental data. In row (A), the model is fit to bivalent and trivalent data simultaneously. In row (B), the model is fit to the bivalent data (solid lines), and the trivalent model with bivalent fit parameters is compared to trivalent data (dashed lines). In row (C), the model is fit to the trivalent data (solid lines), and the bivalent model with trivalent fit parameters is compared to bivalent data (dashed lines). Each panel contains the average root mean squared difference ($\overline{\textrm{RMS}}$) between the model and the average of the experimental data. The rightmost column shows the ratio between the fit base pairing parameter values and the expected base pairing parameter values~\cite{santalucia1998unified}.}
  \label{fgr:Ncmax1_fits}
\end{figure}

\begin{figure}[h]
\centering
  \includegraphics[width=\columnwidth]{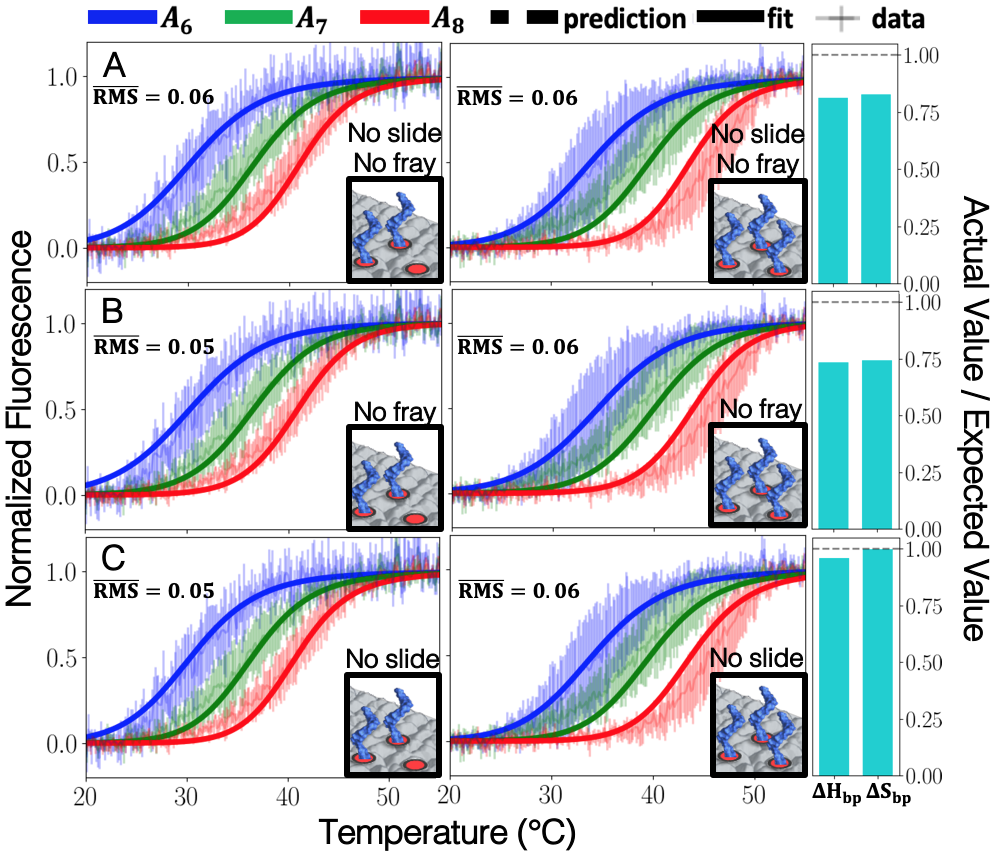}
  \caption{Model with $N_{\mathrm{c,max}}=2$ with exclusion of fraying and/or sliding fit to experimental bivalent and trivalent data simultaneously. In row (A), both fraying and sliding are disallowed. In row (B), sliding is allowed, but fraying is disallowed. In row (C), fraying is allowed, but sliding is disallowed. Each panel contains the average root mean squared difference ($\overline{\textrm{RMS}}$) between the model and the average of the experimental data. The rightmost column shows the ratio between the fit base pairing parameter values and the expected base pairing parameter values~\cite{santalucia1998unified}.}
  \label{fgr:no_slide_no_fray}
\end{figure}

\begin{figure}[h]
\centering
  \includegraphics[width=\columnwidth]{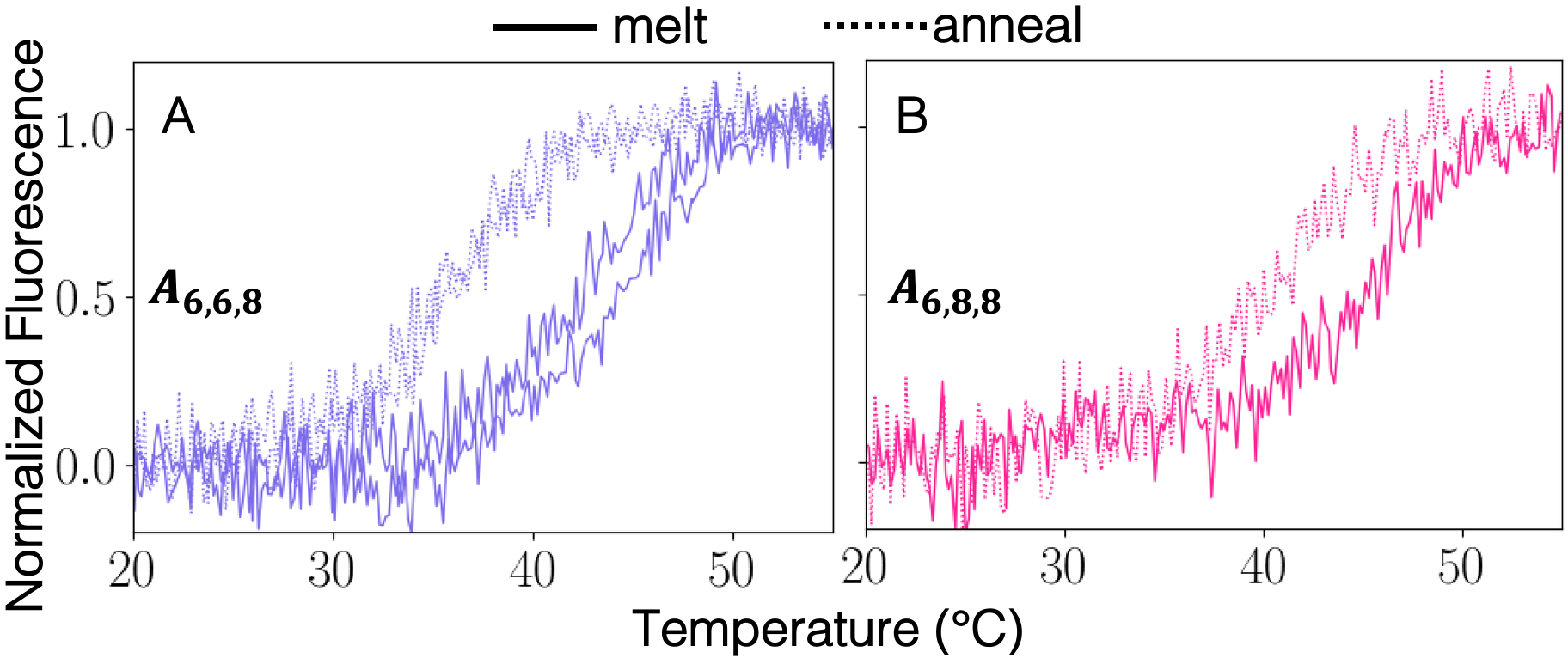}
  \caption{Comparison of melting and annealing curves for mixed hinges. The left panel shows melting (solid) and annealing (dotted) curves for two experimental replicates of a nano-hinge with two 6-base overhangs and one 8-base overhang, and the right panel shows melting (solid) and annealing (dotted) curves for one experimental replicate of a nano-hinge with one 6-base overhang and two 8-base overhangs.}
  \label{fgr:mixed_melt_and_anneal_data}
  
\end{figure}




\end{document}